\documentclass[12pt,dvips]{article}
\usepackage{epsfig}
\newcommand{\acc}{\ensuremath{\mathrm{acc}}}
\newcommand{\boltz}[1]{\ensuremath{\exp({-\mathrm{\beta}#1})}}
\newcommand{\boltzm}[1]{\ensuremath{\exp[{-\mathrm{\beta}#1}]}}
\newcommand{\boltzb}[1]{\ensuremath{\exp\{{-\mathrm{\beta}#1}\}}}
\newcommand{\drv}[1]{\ensuremath{d\mathbf{r}\mathrm{_{#1}}}}
\newcommand{\dyv}[1]{\ensuremath{d\mathbf{y}\mathrm{_{#1}}}}
\newcommand{\drs}[1]{\ensuremath{d\mathbf{r}\mathrm{^{#1}}}}
\newcommand{\rv}[1]{\ensuremath{\mathbf{r}_{#1}}}
\newcommand{\xv}[1]{\ensuremath{\mathbf{x}_{#1}}}
\newcommand{\yv}[1]{\ensuremath{\mathbf{y}_{#1}}}
\newcommand{\uv}[1]{\ensuremath{\mathbf{\hat{u}}_{#1}}}
\newcommand{\vv}[1]{\ensuremath{\mathbf{\hat{v}}_{#1}}}
\newcommand{\wv}[1]{\ensuremath{\mathbf{\hat{w}}_{#1}}}
\newcommand{\xuv}[1]{\ensuremath{\mathbf{\hat{x}}_{#1}}}
\newcommand{\yuv}[1]{\ensuremath{\mathbf{\hat{y}}_{#1}}}
\newcommand{\zuv}[1]{\ensuremath{\mathbf{\hat{z}}_{#1}}}

\newcommand{\fph}[1]{\ensuremath{f(\phi_{#1})}}

\newcommand{\omg}[1]{\ensuremath{\omega_{#1}}}
\newcommand{\ph}[1]{\ensuremath{\phi_{#1}}}

\newcommand{\thet}[1]{\ensuremath{\theta_{#1}}}
\newcommand{\tp}[1]{\ensuremath{{#1}^\top}}
\newcommand{\C}{\ensuremath{\mathrm{C}}}

\newcommand{\Gm}{\ensuremath{\mathrm{\Gamma}}}
\newcommand{\G}[2]{\ensuremath{\mathrm{G}_{#2}(#1)}}

\newcommand{\Lc}[1]{\ensuremath{\mathrm{L}_\mathit{#1}}}
\newcommand{\Omg}[1]{\ensuremath{\mbox{\boldmath $\Omega$}_{#1}}}
\newcommand{\Ph}[1]{\ensuremath{\mbox{\boldmath $\Phi$}_{#1}}}
\newcommand{\Lb}[1]{\ensuremath{\Lambda_{#1}}}
\newcommand{\Trx}[2]{\ensuremath{\mathrm{T}_{#1}^{\mathrm{#2}}}}

\newcommand{\Mpx}[1]{\ensuremath{\mathrm{M'_{#1}}}}
\newcommand{\Mx}[1]{\ensuremath{\mathrm{M_{#1}}}}

\newcommand{\W}[2]{\ensuremath{\mathrm{W}_{#1}^{\mathrm{#2}}}}
\newcommand{\Nst}{\ensuremath{\mathrm{(n)}}}
\newcommand{\Ost}{\ensuremath{\mathrm{(o)}}}

\renewcommand{\baselinestretch}{1.0}

\title{Efficient Monte Carlo Methods for Cyclic Peptides}
\author{Minghong G. Wu and Michael W.\ Deem\\
Chemical Engineering Department\\
University of California\\
Los Angeles, CA\ \ 90095-1592}
\begin{document}
\maketitle
\centerline{\textbf{Abstract}}
\begin{quote}
We present a new, biased Monte Carlo scheme for simulating
complex, cyclic peptides.  Backbone atoms are equilibrated 
with a biased rebridging scheme, and side-chain atoms are
equilibrated with a look-ahead configurational bias Monte Carlo.
Parallel tempering is shown to be an important ingredient
in the construction of an efficient approach.
\end{quote}
\vspace{2in}
{\em Molecular Physics}, to appear.
\newpage
\section{Introduction}
\label{sec-intro}
%importance of peptides

Peptides are of fundamental importance in biological systems. They
regulate homeostasis, particularly thirst, feeding and
pain~\cite{Kandel}, serve as important signaling molecules in the
nervous system~\cite{Li98}, and are used as a chemical defense
mechanism by some organisms~\cite{Olivera}. Peptides have been used
within the biotechnology industry to identify antagonists blocking
various abnormal enzymatic reactions or ligand-receptor
interactions~\cite{Clackson}.  Cyclic peptides or constrained peptides
are often preferred for this application, since such molecules lose
less configurational entropy upon binding~\cite{Alberg}. Cyclic
peptides have backbones with a cyclic topology that is formed either
by the condensation of two sulfhydryl (-SH) groups from two cysteine
side chains or by the dyhydration of the head NH$_2$ and tail
COOH groups.  A classic example of using a cyclic peptide as an
antagonist is the blocking of platelet aggregation by RGD
peptides. The GPIIb/IIIa-fibronectin interaction is known to be
responsible for blood platelet aggregation~\cite{Ruoslahti}. Roughly
eight cyclic peptides of the form CRGDxxxC\put(-5,15){\line(-1,0){53}}
\put(-5,10){\line(0,1){5}} \put(-58,10){\line(0,1){5}} ,
CxxxRGDC\put(-5,15){\line(-1,0){53}} \put(-5,10){\line(0,1){5}}
\put(-58,10){\line(0,1){5}} , and CxxxKGDC\put(-5,15){\line(-1,0){53}}
\put(-5,10){\line(0,1){5}} \put(-58,10){\line(0,1){5}}~that are
effective platelet aggregation blockers were
identified~\cite{II_ONeil}.  Several companies are now pursuing
organic analogs of these RGD peptides in clinical trials.
%difficulty of modeling cyclic peptides
%barriers Simulations with
Although no successful drug has yet been designed by purely
computational methods, the discovery of the RGD peptide and roughly
thirty other pharmaceuticals has benefited in some way from computer
simulation~\cite{Boyd}.

Simulation of complex biomolecules with standard Metropolis Monte
Carlo or conventional molecular dynamics, however, often fails to
sample conformations from the correct Boltzmann distribution. The
difficulty lies in the intrinsic high energy barriers between the
conformations adopted at room- or body-temperature, barriers that
cannot be overcome with these methods.
%other methods not Boltzmann
High temperature~\cite{Bruccoleri} or potential-scaled
\cite{Tsujishita} molecular dynamics can cross these barriers, but
these methods sample from a distribution that is not the one of
interest.

%New Monte Carlo methods useful
The configurational bias Monte Carlo method (CBMC), first developed by
Frenkel, Smit, and de Pablo~\cite{Frenkel,dePablo1}, successfully
samples complex energy landscapes by using local information when
proposing moves.
%why as regards Metropolis MC or MD
%biasing
This method has been successfully applied to long chain molecules~\cite{Smit2}, phase behavior of long chain alkanes~\cite{Smit3,
dePablo2}, and conformations of hydrocarbons within zeolite channels~\cite{Bates}. 
A combination with a generalized concerted rotation scheme,
inspired by the method for alkane chains~\cite{II_Dodd},
has been applied to the simulation of linear and
cyclic peptides~\cite{I_Deem1}.  This approach proved to be especially
efficient in sampling cyclic peptides with barrier-separated
conformations, even when the location of the conformation and energy
barriers were not known {\em a priori}.  For cyclic peptides, this
method changes conformations locally by perturbing backbone 
segments of two to three amino acids. 
Such moves have the
potential to equilibrate large molecules with complex topologies.

Despite the successes of the configurational bias and concerted
rotation scheme, difficulties still remain for complex cyclic
peptides. Long or bulky side chains are not well equilibrated, for
example. In some cases, the backbone of cyclic peptides is not sampled
efficiently, due to the unpredictable presence of large barriers in
the multidimensional torsional-angle free-energy landscape. We
present an integrated methodology for simulation of cyclic peptides.
%Efficiency
Special concerns are given to optimizing and quantifying the
efficiency of our method.  We propose a peptide rebridging scheme,
inspired by a method proposed for
polymers~\cite{Boone,Pant95,Mavrantzas98}, and suitable for backbone
equilibration of peptides. Eight torsional degrees of freedom are
altered with this backbone move.  Implementation of the move is
reduced in all cases to the solution of a one-dimensional numerical
problem. Four approaches to biasing the rebridging moves are proposed
and compared.  For side chain regrowth, we propose two new methods,
`semi-look-ahead' and `look-ahead', inspired by Meirovitch's lattice
scanning method \cite{Meirovitch}. We find that both methods
equilibrate side chains rapidly.  We compare their efficiency and
discuss optimal parameter values.

%parallel tempering
For the most complex cyclic peptides, biased Monte Carlo is still not
optimally efficient. To overcome the remaining barriers to effective
sampling, we add parallel tempering to our range of techniques.
Parallel tempering is a rigorous Monte Carlo method, first proposed
for the study of glassy systems with large free energy
barriers~\cite{geyer91}.  This method has been successfully applied to
spin glasses~\cite{hukushima96,marinari98}, self-avoiding random
walks~\cite{tesi96}, lattice QCD~\cite{Boyd98}, linear
peptides~\cite{Hansmann}, and crystal structure
determination~\cite{Falcioni}.  In parallel tempering, we consider a
set of identical systems, each at a distinct temperature. Each system
is equilibrated with both updating and swapping moves. The swapping
moves couple the systems in such a way that the lowest temperature
system is able to escape from local energy minima without explicit
knowledge of the barriers. This method achieves rigorously correct
canonical sampling, and it significantly reduces the equilibration
time.  We show that the combination of biased Monte Carlo and parallel
tempering achieves effective sampling, quickly overcoming energy
barriers and approaching the Boltzmann distribution.
%both better when we have energy barriers that are not obvious a
%priori
%improving side chain simulation
 
%organization of paper
We define our all-atom, molecular model of peptides in
Sec.~\ref{sec-moleint}.  The peptide rebridging scheme is described in
Sec.~\ref{sec-RBsch}, where technical details are provided.  This
section can be skipped on a first reading, as it is simply an
extension of the method in ref.~\cite{I_Deem1} to include
pre-screening. How biasing can be done is discussed in
Sec.~\ref{sec-biasRB}. In Sec.~\ref{sec-para} we apply the concept of
parallel tempering to our system.  The `semi-look-ahead' and
`look-ahead' methods for side chains are presented in
Sec.~\ref{sec-SLA} and Sec.~\ref{sec-LA}, respectively.  Results for
the simulation of complex, cyclic peptides are given in
Sec.~\ref{sec-results}, where the efficiency of our approach is
demonstrated. We discuss the results in Sec.~\ref{sec-discuss}, and
make our conclusions in Sec.~\ref{sec-conclude}.

%%%%%%%%%%%%%%%%%%%%%%%%%%%%%%%%%%%%%%%%%%%%%%%%%%%%%%%%%%%%%%%%%%%%

\section{Simulation Methods}
\label{sec-simuM}
\subsection{Molecular Model}
\label{sec-moleint}
%forcefield used
%  AMBER, ECEPP, CHARMM, ...
We chose to
use the AMBER force field~\cite{Weiner} with explicit atoms.  Other
suitable potential models are ECEPP~\cite{Kang96} and
CHARMm~\cite{Mackerell}.
%dielectric for water
Dielectric theory was used to estimate solvent effects~\cite{Smith}.
Fast coordinates such as bond lengths and bond angles were fixed at
their equilibrium value.  Only the biologically-relevant, torsional
degrees of freedom were sampled. Nonetheless, this method can be
easily generalized to flexible systems.
%Definition of rigid unit
With this assumption, a molecule is comprised of a set of so-called
`rigid units'. Following the definition in ref.~\cite{I_Deem1}, a rigid
unit consists of a set of atoms and bonds that form a rigid body. The
relative distance between any pair of atoms within a rigid unit is
constant. Adjacent rigid units are connected by a single sigma bond.

The rigid units are labeled from the head NH$_2$ to the tail COOH
group of the peptide.  Each rigid unit has exactly one incoming bond
that starts from the previous unit and ends within it.  All other
bonds that leave the unit are defined to be outgoing bonds.  For
example, a $\mathrm{C_{\alpha}H}$ unit has two outgoing bonds, the
first going to the residue and the second going to the next backbone
unit.  For unit $i$, we define $\theta_i$ to be
 the angle formed by the incoming bond
and the outgoing bond to the next backbone unit.  The
atom that ends the incoming bond is defined to be a head atom, and the
atom that starts the outgoing bond is defined to be a tail atom. We
define \rv{i\mathrm{h}} and \rv{i\mathrm{t}} to be the positions of
the head and tail atoms of unit $i$, respectively.

Rigid units that appear in the backbone are divided
into two topological types. Type A includes all rigid units with identical head
and tail atoms.  Type B includes the CONH amide group, which has
$\thet{i}=0$.
% add torsional angle representations to this figure
Figure~\ref{fig:clsunit} illustrates the geometry of these two types
and the definitions of \thet{i\mathrm{h}} and \thet{i\mathrm{t}}, which are
the angles spanned by $\rv{i\mathrm{t}}-\rv{i\mathrm{h}}$ and
the incoming and outgoing bonds, respectively.

\subsection{Rebridging Scheme}
\label{sec-RBsch}
We display in
Fig.~\ref{fig:CNWKRGDC} a typical cyclic peptide, CNWKRGDC\put(-5,15){\line(-1,0){65}}
\put(-5,10){\line(0,1){5}} \put(-70,10){\line(0,1){5}}.  Although the
chemical functionality of peptides lies mostly in the freely-rotating
side chains, backbone equilibration is important since the backbone
serves as a scaffold for the side chains.  We, therefore, use two
types of biased Monte Carlo moves, chosen at random: movement of a
random segment of the backbone with rigid rotation of the associated side
chains and regrowth of a randomly picked side chain. Here we describe
the backbone move, a peptide rebridging scheme.

%RB scheme
The peptide rebridging scheme is inspired by the concerted rotation~\cite{II_Dodd} and rebridging~\cite{Boone} moves for alkane chains and
the extension of concerted rotation to peptides~\cite{I_Deem1}. Peptide rebridging
causes a local conformational change within the molecule, leaving
the rest of the molecule fixed. Rebridging moves are not only
suitable for cyclic peptides but also suitable for the internal parts
of larger linear peptides and proteins.  The main features that
distinguish our rebridging scheme are the pre-screening process, more
degrees of freedom per move, and more efficient biasing.  We proposed
five variations of rebridging moves, differing in the probabilities of
choosing one of the many possible geometric solutions. They are Metropolis (MT), no
Jacobian (NJ), with Jacobian (WJ), with Jacobian and old solutions
(WJO), and with Jacobian and multiple rotations (WJM). Here we
describe WJ. Other variations will be described in
Section~\ref{sec-biasRB}.  Peptide rebridging is carried out in several
steps:
\begin{enumerate}
\item 
Randomly select two torsional degrees of freedom that are separated by
six other torsional degrees of freedom. We label the two torsional angles as \ph{0} and
\ph{7}.  The eight rigid units, including both ends, are labeled from
unit $0$ to unit $7$.  Backbone positions are denoted \rv{ia},
where $i=0,\ldots,6$ and $a=\mathrm{h}$ (head) or $\mathrm{t}$ (tail).
Figure~\ref{fig:babaa} depicts a segment that is selected to be
rebridged.  
\item 
The angles \ph{0} and \ph{7} are rotated, causing the rigid
units between 0 and 6 to change while leaving the rigid
units before 0 and after 6 unchanged.  The range of rotation is within
$\pm\Delta\ph{\mathrm{max}}$. The two rotations break the
connectivity of the molecule and provide new trial positions for
\rv{1\mathrm{h}} and \rv{5\mathrm{t}}.  We denote the new values by
$\ph{0}'$ and $\ph{7}'$.
\item 
Find all geometrical solutions, $\ph{0}',\ldots,\ph{7}'$, that
re-insert the
backbone units in a valid way between rigid units 1 to 6. How we solve
this geometrical problem will be described below.  If no solution is
found, this move is rejected. Otherwise, calculate the Rosenbluth
factor, $\W{}{\Nst}$, which is defined as
\begin{eqnarray}
\label{eqn:rosen_WJn}
	\W{}{\Nst} &=&
	\sum_{i=1}^{k^\Nst}
	\mathrm{J}_i^\Nst\boltz{\mathrm{U}_i^\Nst}\ ,
\end{eqnarray}
where $k^\Nst$ is the number of geometrical solutions found.
The Jacobian associated with the constraints for the $i$th solution is
$\mathrm{J}_i^\Nst$.
\item 
Pick a solution from these $k^\Nst$ solutions with probability
\begin{equation}
\label{eqn:pick1}
p_{i} = \frac{\mathrm{J}_i^\Nst \boltz{\mathrm{U}_i^\Nst}}{\W{i}{\Nst}}\ .
\end{equation}
\item 
Solve the geometrical problem corresponding to \ph{0} and \ph{7}.
These solutions include the old configuration $\ph{0},\ldots,\ph{7}$
and are used to calculate the old Rosenbluth factor
\begin{eqnarray}
\label{eqn:rosen_WJo}
	\W{}{\Ost}& = &\sum_{i=1}^{k^\Ost}\mathrm{J}_i^\Ost 
	\boltz{\mathrm{U}_i^\Ost}\ ,
\end{eqnarray}
where $k^\Ost$ is the number of solutions in the old geometry.
\item The attempted move is accepted with the probability
\begin{equation}
\label{eqn:acc-WJ}
	\acc(\mathrm{o\rightarrow n}) = \min \left(1,
	\frac{\W{}{\Nst}}{\W{}{\Ost}}\right)\ .
\end{equation}
\end{enumerate}

The Jacobian in eqs.~(\ref{eqn:rosen_WJn}) and
(\ref{eqn:rosen_WJo}) accounts for the fact that when we solve for
the angles $\ph{1},\ldots,\ph{6}$, we do not produce uniform distributions.
The Jacobian is defined by
\begin{eqnarray}
\label{eqn:jac_RB}
\mathrm{J}\left(\frac{\ph{1},\ph{2},\ph{3},\ph{4},\ph{5},\ph{6}}
	{\rv{5\mathrm{t}},\ \uv{6},\ \gamma_6}\right)
& = &    \frac{\uv{6}\cdot\hat{\mathbf{e}}_{3}}{\det|\mathrm{B}|}\nonumber \\
\mathrm{B}_{ij} & = & 
			[\uv{j}\times(\rv{5\mathrm{t}}-\rv{\mathit{j}\mathrm{h}})]_i  \mbox{,
	if  $j\leq3$}, \nonumber  \\
& &                      [\uv{j}\times\uv{6}]_{j-3} \mbox{,  if $j=4,\
	5$} .
\end{eqnarray}
Here \uv{i} is the unit vector of the $i$th incoming bond, and
$\hat{\mathbf{e}}_3$ is a unit vector along the laboratory $z$-axis.  The
Eulerian angle ${\gamma_6}$ is the azimuthal angle of $\hat {\bf
u}_7$ in a spherical coordinate system defined with $\hat {\bf u}_6$
as the $z$-axis.  The angle is measured with respect to the plane
defined by $\hat {\bf u}_6$ and $\hat{\mathbf{e}}_3$.  It is
worth mentioning that in refs.~\cite{I_Deem1} and~\cite{II_Dodd}, the
Jacobian lacked the $\uv{6}\cdot\hat{\mathbf{e}}_{3}$ term. The
Jacobian should be invariant under orthogonal transformations, but the
Jacobians in refs.~\cite{I_Deem1} and~\cite{II_Dodd} are not.  Despite this,
proper sampling was attained, because the omitted terms cancel in
the acceptance ratio.  This cancellation does not occur in rebridging,
since \uv{6} is changed by the rotation of \ph{7}.  The Jacobian
appears as a consequence of the end-atom constraints in the canonical
partition function of a constrained or cyclic molecule.  The Jacobian
is derived in Appendix~\ref{AppA}, where a careful discussion of the
cyclic constraint is given as well.

%Flory's frame
The geometrical problem in rebridging is solved by seeking conserved
quantities. It is conceptually helpful to imagine a
break point in the segment to be regrown.  The rigid units before the
break point are built upon the positions of the preceding units, whereas
the rigid units after the break point are built upon the positions of the following
units.  When \rv{i\mathrm{h}} and \rv{(i-1)\mathrm{t}} are expressed in
local coordinates of the (i-1)th unit, the positions are said to be
defined in `forward notation'.  When we build up these positions from
the opposite direction, the positions are said to be defined in
`backward notation'.  How we choose the break
point depends on the identity of the rigid units to be regrown. 
Rigid units before the break point are always defined by
forward notation and rigid units after the break point are always
defined by backward notation. With forward notation we have
$\rv{1\mathrm{t}}=\rv{1\mathrm{t}}(\ph{1})$,
$\rv{2\mathrm{h}}=\rv{2\mathrm{h}}(\ph{1})$,
$\rv{2\mathrm{t}}=\rv{2\mathrm{t}}(\ph{1},\ph{2})$,
$\rv{3\mathrm{h}}=\rv{3\mathrm{h}}(\ph{1},\ph{2})$, and so on.  With
backward notation, we have
$\rv{5\mathrm{h}}=\rv{5\mathrm{h}}(\ph{6})$,
$\rv{4\mathrm{t}}=\rv{4\mathrm{t}}(\ph{6})$,
$\rv{4\mathrm{h}}=\rv{4\mathrm{h}}(\ph{6},\ph{5})$,
$\rv{3\mathrm{t}}=\rv{3\mathrm{t}}(\ph{6},\ph{5})$, and so on.

We use a variant of Flory's local coordinate system~\cite{Flory69}.
The system was modified for units with $\thet{i}=0$ to reduce the
number of
variables appearing in the constraint equations.
The general formulas for \rv{(i+1)\mathrm{h}}$(\ph{1},\ldots,\ph{i})$ and
\rv{i\mathrm{t}}$(\ph{1},\ldots,\ph{i})$ are
\begin{eqnarray}
\label{eqn:loctrni}
\rv{(i+1)\mathrm{h}}(\ph{1},\ldots,\ph{i}) & = &
	\rv{i\mathrm{t}}(\ph{1},\ldots,\ph{i}) +
	l_{i\mathrm{t},(i+1)\mathrm{h}} \Trx{i}{lab}\Lb{i}\Ph{i}
	\nonumber \\ 
\rv{i\mathrm{t}}(\ph{1},\ldots,\ph{i}) & = &
	\rv{i\mathrm{h}} +
	l_{i\mathrm{h},i\mathrm{t}}\Trx{i}{lab}\Lb{i\mathrm{h}}\Ph{i}
\end{eqnarray}
where
\begin{eqnarray}        
\Lb{i} & \equiv & \left(\begin{array}{ccc} 
		\cos\thet{i}    & 0             & 0 \\ 
		0               & \sin\thet{i}  & 0 \\ 
		0               & 0             & \sin\thet{i}
		\end{array} \right) \nonumber \\ 
\Lb{i\mathrm{h}} &\equiv &\left
		( \begin{array}{ccc} 
		\cos\thet{i\mathrm{h}} & 0              & 0 \\ 
		0               &       \sin\thet{i\mathrm{h}}& 0           \\ 
		0               &       0& \sin\thet{i\mathrm{h}}
		\end{array} \right) \nonumber \\ 
\Ph{i} & \equiv &
	\left( \begin{array}{c} 1 \\ \cos\ph{i} \\ \sin\ph{i}
	\end{array} \right)
\end{eqnarray}
Here $l_{i a,j b}$ denotes the constant distance
between \rv{i a} and \rv{j b}.  We call unit $i$ the
reference unit of unit $i+1$. We use the form of
eq.~(\ref{eqn:loctrni}) because it explicitly isolates the terms
involving the variable \ph{i}.

The labels $ a$ and $ b$ can be either $\mathrm{h}$ or
$\mathrm{t}$. This notation is dropped when the
unit is an A unit. For example, $l_{1,2\mathrm{h}}$ says unit 1 is an A
unit and defines the distance between \rv{1\mathrm{h}}
and \rv{2\mathrm{h}}. In this case we also drop the
head or tail notation for vectors and write 
$\rv{1}=\rv{1\mathrm{h}}=\rv{1\mathrm{t}}$.  

The
transformation from local coordinates to the laboratory coordinates in
forward notation is
\begin{eqnarray}
\Trx{1}{lab} &\equiv & (\begin{array}{ccc} \uv{1} & \vv{1} & \wv{1} \end{array}) 
						\nonumber \\
\Trx{i}{lab}(\ph{1},\ldots,\ph{i-1})& 
	\equiv  & (\begin{array}{ccc} \uv{i} & \vv{i} & \wv{i} \end{array}) \nonumber \\
	& =     &\Trx{1}{lab}\Trx{1\phi}{}\Trx{1\theta}{}\Trx{2\phi}{}\cdots
		\Trx{(i-1)\phi}{}\Trx{(i-1)\theta}{} 
		\nonumber  \\
\Trx{i\theta}{} & \equiv&\left( \begin{array}{ccc}
				\cos\thet{i} & -\sin\thet{i} & 0 \\
				\sin\thet{i} & \cos\thet{i} & 0 \\      
				0       &       0       &     1
				\end{array} 
			\right)         \nonumber \\
\Trx{i\phi}{} & \equiv &\left\{ \begin{array}{ll}
				\left(  \begin{array}{ccc}
					1       &       0       &    0          \\
					0       &\cos\ph{i}     & -\sin\ph{i}   \\
					0       &\sin\ph{i}     & \cos\ph{i}            
					\end{array} 
				\right) & \mbox{, if } \thet{i}\neq0 \\
				\left(  \begin{array}{ccc}
					1        & 0    & 0 \\
					0        & 1    & 0 \\  
					0        & 0    & 1 
					\end{array} 
				\right) & \mbox{, if } \thet{i}=0\ .
			  \end{array}
		  \right.
\end{eqnarray}
Here \uv{i}, \vv{i}, and \wv{i} are the axes of the local coordinates
of unit $i$ in forward notation
in the laboratory frame.  The last matrix is modified from
Flory's coordinate system. It is defined so that
$\Trx{i}{lab}=\Trx{i-1}{lab}$ when $\thet{i}=0$. This definition
simplifies our algorithm.

For the rigid units beyond the break point, we use backward
notation. In backward notation, unit $i+1$ is the reference unit of
unit $i$. The general formulas for
\rv{(i-1)\mathrm{t}}$(\ph{6},\ldots,\ph{i+1})$ and
\rv{i\mathrm{h}}$(\ph{6},\ldots,\ph{i+1})$ are
\begin{eqnarray}
\label{eqn:locrbi}
\rv{(i-1)\mathrm{t}}(\ph{6},\ldots,\ph{i+1})    & =     & 
	\rv{i\mathrm{h}}(\ph{6},\ldots,\ph{i+1}) + 
	l_{(i-1)\mathrm{t},i\mathrm{h}}\Trx{i}{lab}\Lb{i}\Ph{i+1} \nonumber  \\
\rv{i\mathrm{h}}(\ph{6},\ldots,\ph{i+1})        & =     & 
	\rv{i\mathrm{t}} + l_{i\mathrm{h},i\mathrm{t}}\Trx{i}{lab}\Lb{i\mathrm{t}}\Ph{i+1}
		\nonumber \\
\Lb{i\mathrm{t}}&\equiv &\left( \begin{array}{ccc}
			\cos\thet{i\mathrm{t}}   & 0            & 0 \\
			0        & \sin\thet{i\mathrm{t}}       & 0             \\      
			0        & 0                    & \sin\thet{i\mathrm{t}} 
			\end{array} 
		\right)\ .      
\end{eqnarray}
The transformation from local coordinates
to the laboratory coordinates in backward notation is
\begin{eqnarray}
\Trx{i}{lab}    & \equiv& (\begin{array}{ccc} \xuv{i} & \yuv{i} & \zuv{i} \end{array}) 
		\nonumber \\
& =     & \Trx{5}{lab}\Trx{6\phi}{}\Trx{5\theta}{}\Trx{5\phi}{}\cdots
	\Trx{(i+2)\phi}{}\Trx{(i+1)\theta}{} \ .
\end{eqnarray}
Here \xuv{i}, \yuv{i}, and \zuv{i} are the axes of the local coordinates
of unit $i$ in backward notation
in the laboratory frame.
%Solving the geometrical problem
%rebridging
%solution of equations numerically

For all rebridging cases that we consider, it is possible to find
three constraint equations with three independent torsional angles and
to determine the solutions by solving a one-dimensional equation
numerically. The constraint equations vary depending on the types of
the units 1 to 5.  Table~\ref{tab:0pro} lists the six distinct cases
that can occur and the corresponding constraint equations.  The
dependencies of the backbone positions on the torsional angles are
specified explicitly in the argument, and one can tell from the
arguments if the positions are in forward or backward
notation. Actually, cases 4 and 5 are mirror images of case 1 and 2,
with the rigid units labeled in the opposite direction.  Strictly
speaking, then, there are only four distinct cases: cases 1, 2, 3, and 6.

%simplified tables
In all cases, the first two constraint equations have at most two
independent variables each.  In one special case (case 3), an equation
with only one variable is found.  In the first three cases, the first
two equations of each set are used to derive two torsional
angles as analytic functions of \ph{1}.  These two analytic
expressions are in turn substituted into the third equation, which is
solved numerically in the \ph{1} domain.  In cases 1 and 2 the other
two independent angles are \ph{2} and \ph{6}. Case 3 is special
because \ph{2} is a constant. The other torsional angle needed in the
third constraint equation is \ph{6}.  In cases 4 and 5, the approach is
similar, except that the equations are solved numerically in the \ph{6}
domain. In case 6, the second and fourth rigid units are
arbitrary and can be either A or B.  This is a special case
in which \rv{3} can be written as a function of a single, new
torsional angle.  This case includes, for example, ABABA, which
corresponds to
$\mathrm{C}_\alpha$-amide-$\mathrm{C}_\alpha$-\mbox{amide}-$
\mathrm{C}_\alpha$. It is obvious that the geometrical
constraints keep the distances $|\rv{1}-\rv{3}|$ and $|\rv{3}-\rv{5}|$
constant in all possible solutions. We also know the trial distance
$|\rv{1}-\rv{5}|$ after performing the two rotations.  These distances
are conserved in all solutions.  Therefore, possible positions
of \rv{3} should fall on the intersection of two spheres centered at
\rv{1} and \rv{5}. Figure~\ref{fig:ababa} shows the geometry of this
segment and the conserved distances. If the triangle inequality
\begin{equation}
\label{eqn:triieq}
l_{1,3} + l_{3,5} \geq |\rv{1}-\rv{5}|_\mathrm{trial}
\end{equation}
holds, we can define a new set of local coordinates for unit 3:
\begin{eqnarray}
\uv{3}' &=      & (\rv{5}-\rv{1})/l_{1,5}
\nonumber \label{eqn:locnew}\\
\wv{3}' &=      &\uv{1}\times\uv{3}'/|\uv{1}\times\uv{3}'| \nonumber \\
\vv{3}' &=      &\wv{3}'\times\uv{3}'
\end{eqnarray}
to obtain an expression for \rv{3} as a function of a single, new angle $\ph{3}'$:
\begin{eqnarray}
\label{eqn:locmtxn}
\rv{3}(\ph{3}') &=      &\rv{1}+l_{1,3}{\Trx{3}{lab}}'\Lb{3}'\Ph{3}'\ ,
\end{eqnarray}
where
\begin{eqnarray}
{\Trx{3}{lab}}' & \equiv & (\begin{array}{ccc} \uv{3}' & \vv{3}' & \wv{3}' \end{array}) 
	\nonumber \\
\Lb{3}' & \equiv & \left(\begin{array}{ccc}
			\cos\thet{3}' & 0        & 0 \\
			0        & \sin\thet{3}' & 0 \\ 
			0       &       0       & \sin\thet{3}' 
		    \end{array} \right) \nonumber \\
\thet{3}' &\equiv       & \left|\cos^{-1}
	\frac{{l_{1,3}}^2+{l_{1,5}}^2-{l_{3,5}}^2}{2l_{1,3}l_{1,5}}\right|\nonumber \\
\Ph{3}' & = & \left(    \begin{array}{c}
			1 \\
			\cos\ph{3}' \\
			\sin\ph{3}'
			\end{array} 
		\right)\ .
\end{eqnarray}

All of the constraint equations in table~\ref{tab:0pro} can be
grouped by their functional forms into four types. The fourth column of
table~\ref{tab:0pro} shows the type of each constraint equation.  The
general functional forms of these constraint equations are listed in
table~\ref{tab:constype}.  The first type is a quadratic function of a
single variable. The other types are
functions of two torsional angles.  They are based on
either conserved distances, as in `dist', or conserved
angles, as in `dot' and `dot1'.  The last column of
table~\ref{tab:constype} lists the characteristic matrix, which is
used in the pre-screening process and the evaluation of the third target function.
%prescreening--leads to many cases

We illustrate the peptide rebridging algorithm by taking case 6 in table~\ref{tab:0pro}
as an example.  If eq.~(\ref{eqn:triieq}) is not satisfied, the
trial move is immediately rejected because of a geometrical failure.
Otherwise, we go on.
The first constraint equation
allows us to express \ph{1} in terms of $\ph{3}'$. To do this, we rewrite the
constraint equation as
\begin{eqnarray}
0  = [\rv{3}(\ph{3}')-\rv{2\mathrm{h}}]^{\mathrm{\top}}
[\rv{3}(\ph{3}')-\rv{2\mathrm{h}}]-{l_{2\mathrm{h},3}}^2 \nonumber
\end{eqnarray}
and use
eqs.~(\ref{eqn:loctrni}) and eq.~(\ref{eqn:locmtxn}) to obtain
\begin{eqnarray}
\label{eqn:consababa}
0 &= &(l_{1,3}{\Trx{3}{lab}}'\Lb{3}'\Ph{3}'-l_{1,2\mathrm{h}}
	\Trx{1}{lab}\Lb{1}\Ph{1})^{\mathrm{\top}}
(l_{1,3}{\Trx{3}{lab}}'\Lb{3}'\Ph{3}'-l_{1,2\mathrm{h}}
	\Trx{1}{lab}\Lb{1}\Ph{1})-{l_{2\mathrm{h},3}}^2\nonumber \\
 &= &{l_{1,3}}^2+{l_{1,2\mathrm{h}}}^2-{l_{2\mathrm{h},3}}^2-
2l_{1,3}l_{1,2\mathrm{h}}\tp{\Ph{3}'}\tp{\Lb{3}'}\tp{{\Trx{3}{lab}}'}
	\Trx{1}{lab}\Lb{1}\Ph{1}\ .
\end{eqnarray}
We introduce the constant matrix
\begin{equation}
\C\equiv\left( \begin{array}{ccc}
		1       &0      &0 \\
		0       &0      &0 \\
		0       &0      &0 \end{array} \right) 
\end{equation}
and multiply the first three constant terms in
eq.~(\ref{eqn:consababa}) by $\tp{\Ph{3}'}\C\Ph{1}$, which is unity,
to obtain
\begin{equation}
\tp{\Ph{3}'}\Mx{}\Ph{1} =       0\ , \label{eqn:pmpp}
\end{equation}
where the constant characteristic matrix \Mx{} is defined as
\begin{equation}
\label{eqn:Mx_case_6}
\Mx{}=  ({l_{1,3}}^2+{l_{1,2\mathrm{h}}}^2-{l_{2\mathrm{h},3}}^2)\C
	-
2l_{1,3}l_{1,2\mathrm{h}}\tp{\Lb{3}'}\tp{({\Trx{3}{lab}}')}\Trx{1}{lab}\Lb{1}\ .
\end{equation}
In each case, the first two constraint equations can be cast into the form
of eq.~(\ref{eqn:pmpp}).  The right hand column of table~\ref{tab:constype}
lists the constraint equations and the corresponding
characteristic matrix \Mx{} for each
case. Equation~(\ref{eqn:Mx_case_6}), for example, is a
special case of the `dist' type constraint equation in
table~\ref{tab:constype}, in which we have $\rv{i'}=\rv{j'}=\rv{1}$.

To solve the constraint equations, we set $\omg{i}=\cos(\ph{i}/2)$ and use
\begin{eqnarray}
\label{eqn:triden}
\cos\ph{i}&     =&(1-\omg{i}^2)/(1+\omg{i}^2) \nonumber \\
\sin\ph{i}&     =&2\omg{i}/(1+\omg{i}^2)
\end{eqnarray}  
to replace each $\cos\ph{i}$ and $\sin\ph{i}$ in
eq.~(\ref{eqn:pmpp}). We rewrite eq.~(\ref{eqn:Mx_case_6}) as
\begin{eqnarray}
\tp{\Omg{3}'}\Mpx{}\Omg{1} &=   &0\ , \label{eqn:omo}
\end{eqnarray}
where
\begin{eqnarray}
\Omg{i}&\equiv  &\left( \begin{array}{c}
			1 \\
			\omg{i} \\
			\omg{i}^2
			\end{array}
		 \right) \ .
\end{eqnarray}
The matrix \Mpx{} is related to \Mx{} by 
\begin{equation}
\label{eqn:MtoMp}
\Mpx{}  = \left( 
			\begin{array}{ccc}
\Mx{11}+\Mx{12}+\Mx{21}+\Mx{22}\;\;\;\;\;\; & 2(\Mx{13}+\Mx{23})
&\;\;\;\;\;\; \Mx{11}-\Mx{12}+\Mx{21}-\Mx{22}\\
2(\Mx{31}+\Mx{32})                  & 4\Mx{33}             & 2(\Mx{31}-\Mx{32}) \\
\Mx{11}+\Mx{12}-\Mx{21}-\Mx{22}\;\;\;\;\;\; & 2(\Mx{13}-\Mx{23}) 
&\;\;\;\;\;\; \Mx{11}-\Mx{12}-\Mx{21}+\Mx{22}
			
			\end{array}
		\right)\; .
\end{equation}
Equation~(\ref{eqn:omo}) is quadratic in \omg{1}, and we find
\begin{eqnarray}
\label{eqn:quad}
\omg{1} &=      &\frac{1}{2c_2}
	\left[-c_1\pm (c_1^2-4c_0c_2)^{\frac{1}{2}}\right]\nonumber \\
	&=      &f_{1\pm}(\omg{3}')\ ,
\end{eqnarray}
where
\begin{eqnarray}
c_0     &=      &\Mpx{11}+\Mpx{21}\omg{3}'+\Mpx{31}{\omg{3}'}^2 \nonumber \\
c_1     &=      &\Mpx{12}+\Mpx{22}\omg{3}'+\Mpx{32}{\omg{3}'}^2 \nonumber \\
c_2     &=      &\Mpx{13}+\Mpx{23}\omg{3}'+\Mpx{33}{\omg{3}'}^2\; .
\end{eqnarray}
Since $\omg{3}'$ must produce a non-negative discriminant in
eq.~(\ref{eqn:quad}) in order to produce a real-valued \omg{1}, pre-screening can
be done by solving
\begin{equation}
\label{eqn:discri}
c_1^2-4c_0c_2=0 \ .
\end{equation}
Equation~(\ref{eqn:discri}) is a quartic polynomial equation.  We use
an eigenvalue
method to solve this equation and to determine the valid
domains of $\ph{3}'$ in the first constraint
equation~\cite{C_recipes}. Note that $\ph{1} =
f_{1\pm}(\ph{3}')$ has two branches.

Following a derivation parallel to that in 
eqs.~(\ref{eqn:consababa})--(\ref{eqn:discri}), 
we can write $\ph{6}=f_{2\pm}(\ph{3}')$. 
A similar pre-screening process is done to
determine the valid domains of $\ph{3}'$ in the second equation. This
pre-screening process reduces the CPU cost and increases the
efficiency of the algorithm considerably.

Evaluation of the third target function is performed over the valid
domains of $\ph{3}'$, which are the intersections of the valid domains
found by pre-screening.  The independent variable is chosen to be
$\ph{3}'$ instead of $\omg{3}'$, since the latter may
be valid on an infinite domain.  To find the acceptable new rigid
unit positions, the third target function is solved. To evaluate the
target function, a series of calculations is repeated for each
$\ph{3}'$. First, we calculate the corresponding \ph{1} and
\ph{6}. Second, we determine $\rv{3}(\ph{3}')$,
$\rv{2\mathrm{h}}(\ph{1})$, and $\rv{4\mathrm{t}}(\ph{6})$. Third, we
calculate \rv{2\mathrm{t}} and \rv{4\mathrm{h}}, which are uniquely
determined by the trial $\rv{3}(\ph{3}')$, $\rv{2\mathrm{h}}(\ph{1})$,
and $\rv{4\mathrm{t}}(\ph{6})$ 
(see figure~\ref{fig:ababa}).  Finally,
we substitute \rv{2\mathrm{t}} and \rv{4\mathrm{h}} into the target
function.
We evaluate the target function on a grid, using a grid width of 0.003 
radians.  A finer grid
is used when the function approaches zero.  The function values
so obtained are used to locate approximately the roots.  Brent's
method is used to refine the roots~\cite{C_recipes}.  The
roots for $\ph{3}' $ are sufficient to determine all the backbone
positions.  Substituting each root into $f_{1\pm}$ and $f_{2\pm}$, we obtain
\ph{1} and \ph{6}, and thus \rv{2\mathrm{h}}, \rv{4\mathrm{t}}, and
\rv{3}.  Other backbone positions can be calculated easily.
Side chains are rigidly rotated so as to connect to the backbone
properly, and the geometrical problem is solved.

For each valid $\ph{3}'$, there are two branches of the solution for $\ph{1} =
f_{1\pm}(\ph{3}')$ 
and also two branches of the solution for $\ph{6} =
f_{2\pm}(\ph{3}')$.  Therefore, the target function has
four branches.  Figure~\ref{fig:2-2var} shows a typical target
function. In the case shown in figure~\ref{fig:2-2var},
there are six solutions.

In summary, the algorithm for solving the geometrical problem for case 6
works as follows:
\begin{enumerate}
\item
If the geometry does not satisfy eq.~(\ref{eqn:triieq}), the move is rejected.
\item
Calculate the characteristic matrices, \Mx{1} and \Mx{2}, of the first
two constraint equations.  Transform \Mx{1} to \Mpx{1} and \Mx{2} to
\Mpx{2}, using eq.~(\ref{eqn:MtoMp}).  Find the intersection of valid
domains using eq.~(\ref{eqn:discri}).  If no common
domain is found, the move is rejected.
\item
Search for roots of the third equation on the valid $\ph{3}'$ domains.
Determine all the backbone positions associated with each solution for
$\ph{3}'$.  Determine the positions of all associated side chains.
\end{enumerate}
Other cases in table~\ref{tab:0pro} are solved similarly,
except that the independent variable is either \ph{1} or \ph{6}.

\subsection{Biasing of the Rebridging Moves}
\label{sec-biasRB}
%NJ, WJ, WJO, WJM
There are several ways to bias solutions in the rebridging scheme.
The first method is discussed in~\cite{I_Deem1}. The Rosenbluth factors
are defined as
\begin{eqnarray}
\label{eqn:rosenb_CR}
	\W{}{\Nst}& = &\sum_{i=1}^{k^\Nst} \boltz{\mathrm{U}_i^\Nst}
	\nonumber \\
	\W{}{\Ost}& = &\sum_{i=1}^{k^\Ost} \boltz{\mathrm{U}_i^\Ost}
\end{eqnarray}
The proposed move is accepted with the probability
\begin{eqnarray}
\label{eqn:acc-CR}
	\acc(\mathrm{o\rightarrow n}) &= &\min
	\left(1, \frac{\mathrm{J}^\Nst \W{}{\Nst}}{\mathrm{J}^\Ost \W{}{\Ost}}
\right)  \ ,
\end{eqnarray}
where $\mathrm{J}$ is the Jacobian.
This method is called no Jacobian (NJ).

A second method of bias, called with
Jacobian (WJ), includes the bias introduced by the Jacobian within
the Rosenbluth factors, as in eqs.~(\ref{eqn:rosen_WJn}) and
(\ref{eqn:rosen_WJo}).  The proposed move is accepted with the
probability given by eq.~(\ref{eqn:acc-WJ}).  This approach is
expected to achieve better sampling than NJ, since it
explicitly includes the bias introduced by the Jacobian within the
move \cite{Escobedo_95}.

A third method of bias includes the old and new solutions within a single Rosenbluth factor.
Solutions are picked with the probability
\begin{eqnarray}
\label{eqn:pickboth}
p_{i} & = &\frac{\mathrm{J}_i\boltz{\mathrm{U}_i}}{\W{}{}}\; \mbox{, }i=1,\ldots, 
(k^\Ost + k^\Nst) \nonumber \\
\W{}{} & = & \W{}{\Ost}+\W{}{\Nst} \ .
\end{eqnarray}
Here the Rosenbluth factors are, again, defined as in eqs.~(\ref{eqn:rosen_WJn}) and
(\ref{eqn:rosen_WJo}).
Such a move is always accepted, although the new state may
be identical with the old state. This method is called with Jacobian
and old solutions (WJO).

It is possible to perform multiple rotations on \ph{0} and \ph{7}.
This scheme must be based on WJ or NJ so as to satisfy detailed
balance. We choose WJ and call this method with Jacobian and multiple
rotations (WJM).  For a rebridging move with $k_{\max}$ rotations,
$k_{\max}-1$ rotations around the old state must be performed
to obtain a correct old Rosenbluth factor.  A new configuration is
selected from the solutions with the probability
\begin{equation}
\label{eqn:pickWJM}
p_i  = \frac{\mathrm{J}_i^\Nst\boltz{\mathrm{U}_i^\Nst}}
	{\sum_{k=1}^{k_{\max}}\W{k}{\Nst}}\ ,
\end{equation}
where $\W{k}{\Nst}$ is the Rosenbluth factor of the $k$th rotation, as
calculated by eq.~(\ref{eqn:rosen_WJn}). The acceptance probability is
\begin{eqnarray}
\label{eqn:acc-WJM}
	\acc(\mathrm{o\rightarrow n}) &= &\min
	\left(1, \frac{\sum_{k=1}^{k_\mathrm{max}}{\W{}{\Nst}(k)}}
		{\sum_{k'=1}^{k_\mathrm{max}}{\W{}{\Ost}(k')}}\right)\ .
\end{eqnarray}

The last method is based on Metropolis rules, in which a solution is
picked at random without any bias, as in ref.~\cite{II_Dodd}. The
picking probability and acceptance criteria are
\begin{eqnarray}
p_{i} & = &\frac{1}{k^\Nst}\nonumber \\
\acc(\mathrm{o\rightarrow n}) &= &\min \left[1, 
			\frac{\mathrm{J}^\Nst k^\Nst \boltz{\mathrm{U}^\Nst}}
			     {\mathrm{J}^\Ost k^\Ost
\boltz{\mathrm{U}^\Ost}}\right]\ .
\end{eqnarray}
The method is called Metropolis (MT).

%more than one trial angle

%parallel tempering
%definition of method
\subsection{Parallel Tempering}
\label{sec-para}

%partition function
The use of biasing mitigates, but does not eliminate, the various free
energy barriers in cyclic peptides. Even a small cyclic peptide is, in
a sense, a `glassy' system due to these significant and
unpredictably-located free energy barriers. To deal with this issue,
we use parallel tempering~\cite{geyer91}.

In parallel tempering we consider an extended ensemble with $n$
systems, labeled as $i=1,\ldots,n$.  Each system is a copy of the
original system, except that each is equilibrated at a distinct
temperature, $T_i$, where $i=1,\ldots, n$ and $T_1 < T_2 <\ldots <
T_n$. The canonical partition function of this extended canonical
ensemble is given by
\begin{equation}
{\mathrm Q}  = \prod_{i=1}^n {\mathrm Q}_i\ , 
\label{eq:part_pt}
\end{equation}
where ${\mathrm Q}_i$ is the individual canonical partition function
of the $i$th system.  Two types of moves are performed in the
ensemble. The first is a regular Monte Carlo move within a randomly chosen
system.
%swapping move
The second is a swapping move. A swapping move proposes to exchange the 
configurations of the two systems $i$ and $j=i+1$, $1\leq i< n$.
This move is accepted with the probability
\begin{eqnarray}
\label{eqn:acc-pt}
\acc[(i,\ j)\rightarrow (j,\ i)] & = &
	\min[1,\exp(-\beta_i\mathrm{U}_j-\beta_j\mathrm{U}_i+\beta_i\mathrm{U}_i+
		\beta_j\mathrm{U}_j)] \nonumber \\
&               = &\min[1,\exp(-\Delta\beta\Delta\mathrm{U})]\; .
\end{eqnarray}
This technique forces each system to sample the Boltzmann distribution
at the appropriate temperature. In our case, we are interested in the
lowest temperature distribution only. The higher temperature systems
are included solely to help the lowest temperature system to escape
from local energy minima via the swapping moves.  To achieve efficient
sampling, the highest temperature should be such that no significant
free energy barriers are observed. To ensure that the swapping moves
are accepted, the energy histograms of adjacent systems should
overlap.  The sampling efficiency is modestly affected by the fraction
of overlap. We arbitrarily chose to adjust the temperatures so that the
probability of accepting a swapping move was roughly 0.1,
and no attempt was
made to optimize further these temperatures. We will show that the extra
computational cost of simulating the higher temperature systems is
more than compensated for by the increased sampling efficiency of the
lowest temperature system.

%CBMC
\subsection{Semi-Look-Ahead}
\label{sec-SLA}
%semi-look ahead (as compared to bad old method)
Since the conformations of the side chains determine the biological
activity of peptides, effective sampling of side chains is important.
Our method is based on the side chain dihedral angle moves in ref.~\cite{I_Deem1}. A
finite regrowth probability is assigned to each side chain of the
molecule. A side chain move proceeds by regrowing the side chain unit
by unit, beginning from the bond connecting the backbone to the side
chain. At each step, $n_1$ twigs are generated and used to calculate
the new partial Rosenbluth factor.  One of the twigs is selected with
a probability proportional to the Boltzmann factor associated with
that twig. The old configuration and $n_{\alpha}-1$ random twigs are
generated and used to calculate the old partial Rosenbluth
factor. This procedure is repeated until the end of chain is
reached. The new chain is accepted with the probability
\begin{equation}
	\label{eqn:acc1} \acc(\mathrm{o\rightarrow n}) = \min\left(1,
	\frac{\W{}{\Nst}}{\W{}{\Ost}}\right)\ ,
\end{equation}
where \W{}{\Nst} is the product of the new partial Rosenbluth
factors, and \W{}{\Ost} is the product of the old partial Rosenbluth
factors.

We propose a new method called semi-look-ahead (SLA) for side chain
regrowth.  For each torsionally-flexible bond, we define the group of
atoms included in the partial Rosenbluth factor to be the maximum set
of atoms whose positions are uniquely determined by choosing the trial
rotation of this bond.  Figure~\ref{fig:sidechai} sketches this new
definition of atom groups and contrasts it with the one in
ref.~\cite{I_Deem1}.  Our definition includes atoms beyond the
boundary of rigid units, including the head atoms of rigid units
adjacent to the current one.  We expect SLA to achieve better sampling
efficiency and faster equilibration than the method without
look-ahead~\cite{I_Deem1}, due to the improved energy estimate
for the biasing.

The incremental energy at each step can be
split into the internal and external components~\cite{Smit1}. With this
decomposition, torsional angles
are generated with a probability derived from the internal
energy, and the partial Rosenbluth factors include the external energy
only.  In our system, only torsional energies can be put
into internal energy, and these energies account for only a small
fraction of the total energy. We find it most efficient to set the internal energy to
zero and to include all of the energy within the external component.

%look ahead
\subsection{Look Ahead}
\label{sec-LA}
%description
%detailed balance
%intuitive reason for W(n)/W(o)
For long chains or chains with bulky units, a more extensive form of
look-ahead may help to avoid proposing high energy
configurations~\cite{Meirovitch}.  The idea is illustrated in
figure~\ref{fig:lookahead}. If we regrow the molecule by exploring the
energy landscape only one rigid unit ahead, we will choose one
configuration, as in figure~\ref{fig:lookahead}a.  If we can look
ahead two rigid units at one time, we may find the high energy region
associated with 
that configuration and choose a more likely one instead, as in
figure~\ref{fig:lookahead}b.

We proposed two methods for look-ahead.  The idea is to include a
contribution from the energetic surroundings of the succeeding unit within the
Rosenbluth factor.  The first method, look-ahead (LA), generates $n_1$
trial rotations of the unit to be regrown and $n_2$ trial rotations
of the succeeding unit for each of the trial rotations of the first
unit.  In the second method, we set $n_1=n_2=n$. We generate $n$
configurations of the first rigid unit, with $n$ configurations of the
second unit associated to each. When regrowing the second unit, we use the $n$
configurations already proposed during the regrowth of the first
configuration. We, therefore, generate only the configurations for
the third unit when regrowing the second
unit. This method of look-ahead with recycled configurations is abbreviated as LARC.

We now describe the procedure for carrying out these methods.
%LA, LARC
Suppose we want to cut and regrow
rigid units $i=1,\ldots,N$. The following procedure describes
how to generate and accept these units:
\begin{enumerate}
\item \label{itm:beg}Generate a set of $n_1$ trial torsional angles
$\{{\ph{1}(\alpha)}\}$, $\alpha=1,\ldots,n_1$. Each angle is generated
according to the internal potential of unit 1
\begin{eqnarray}
\label{eqn:p_int_1}
p_{1}^\Nst(\alpha)&=&\mathrm{C_1}\boltzb{\mathrm{U_1^{int}}[\ph{1}(\alpha)]}\ .
\end{eqnarray}
Denote the external energy of unit 1 at $\ph{1}(\alpha)$ by
$\mathrm{U_{1}^{ext}}(\alpha)$.
\item 
\label{itm:ahead1}For each trial $\ph{1}(\alpha)$, generate a set of $n_2$ torsional
angles $\{{\ph{2}(\alpha,\ \gamma)}\}$.  Each angle is generated according to the
internal potential
\begin{eqnarray}
p_2^\Nst(\alpha,\ \gamma) &=
&\mathrm{C_2}\boltzb{\mathrm{U_2^{int}}[\ph{2}(\alpha,\ \gamma)]}\ .
\end{eqnarray}
Denote the external energy of unit 2 at $\ph{1}(\alpha)$ and $\ph{2}(\alpha,\ \gamma)$
by $\mathrm{U_2^{ext}}(\alpha,\ \gamma)$. For LA, the number $n_1$ can be
different from $n_2$. For LARC $n_1=n_2$.
\item 
\label{itm:ahead2}Define
\begin{eqnarray}
\mathrm{w}_2^\Nst(\alpha)&= &\frac{\sum_{\gamma
=1}^{n_2}\boltzm{\mathrm{U_2^{ext}}(\alpha,\ \gamma)}}{n_2}
\end{eqnarray}
and
\begin{eqnarray}
\mathrm{w}_1^\Nst&= &\frac{\sum_{\alpha=1}^{n_1}
	\sum_{\gamma=1}^{n_2}\boltzm{\mathrm{U_1^{ext}}(\alpha)}
	\boltzm{\mathrm{U_2^{ext}}(\alpha,\ \gamma)}}
	{n_1n_2}
\nonumber \\
& = & \frac{\sum_{\alpha =1}^{n_1}\boltzm{\mathrm{U_1^{ext}}(\alpha)}
	\mathrm{w}_2^\Nst(\alpha)}{n_1}\ .
\end{eqnarray}
\item 
\label{itm:end} Pick a $\ph{1}(\alpha)$ with the probability
\begin{eqnarray}
\label{eqn:pick_LA}
q_1^\Nst(\alpha)&=&
\frac{\boltzm{\mathrm{U_1^{ext}}(\alpha)}\mathrm{w}_2^\Nst(\alpha)}
	{\mathrm{w}_1^\Nst}\ .
\end{eqnarray} 
To simplify the notation, we switch the labels of the chosen $\alpha$th angle
with the first torsional angle so that the chosen angle is first.
\item 
Repeat steps (\ref{itm:beg})-(\ref{itm:end}) for rigid unit 2 to unit N-1, except
that for LARC, the $n_2$ twigs of unit 2 corresponding to the
chosen unit 1 are recycled to be the $n_1$
($n_1=n_2$) trial configurations of unit 2, and so on.
\item 
For the Nth unit, which is the last unit, there is no need to look
ahead, so we repeat step (1) and generate $n_1$ torsional
angles $\{\ph{N}(\alpha)\}$.  Calculate 
\begin{eqnarray}
\mathrm{w_N^\Nst}
&\equiv&\sum_{\alpha
=1}^{n_1}\boltzm{\mathrm{U_\mathit{N}^{ext}}(\alpha)}\ ,
\end{eqnarray}
and pick a $\ph{N}(\alpha)$ with the probability
\begin{eqnarray}
q_N^\Nst(\alpha)&=& \frac{\boltzm{\mathrm{U_\mathit{N}^{ext}}
	(\alpha)}}{\mathrm{w}_N^\Nst}\ .
\end{eqnarray}
\end{enumerate}
We also need to generate and calculate the old Rosenbluth weights:
\begin{enumerate}
\item 
Generate $n_1-1$ trial torsional angles with the probability given by
eq.~(\ref{eqn:p_int_1}).  These angles and the original angle comprise a
set of torsional angles $\{\ph{1}(\alpha)\}$.  Let the original angle be
labeled as $\ph{1}(1)$.
\item 
For each $\ph{1}(\alpha)$ other than $\ph{1}(1)$, generate $n_2$
torsional angles $\{\ph{2}(\alpha,\ \gamma)\}$.  For the original angle $\ph{1}(1)$,
generate $n_2$ angles if the method is LA and $n_2-1$
angles if the method is LARC. For LARC add the original \ph{2} to the
set of angles generated and label the original angle as $\ph{2}(1,\ 1)$.
All configurations other than the original one are generated according
to the probability
\begin{eqnarray}
p_2^\Ost(\alpha,\
\gamma)&=&\mathrm{C_2}\boltzb{\mathrm{U_2^{int}}[\ph{2}(\alpha,\
\gamma)]}\ .
\end{eqnarray}
Define $\mathrm{w}_2^\Ost(\alpha)$ and $\mathrm{w}_1^\Ost$ in an analogous way as
$\mathrm{w}_2^\Nst(\alpha)$ and $\mathrm{w}_1^\Nst$.
\item 
Repeat the preceding two steps for unit 2 to N-1.
\item 
For the Nth unit, which is the last unit, generate a set of
$n_1-1$ angles $\{\ph{N}(\alpha)\}$ with the probability given
by eq.\ (\ref{eqn:p_int_1})
Add the original angle.  Calculate $\mathrm{w}_N^\Ost$.

\end{enumerate}
The proposed move is accepted with the probability
\begin{eqnarray}
\label{eqn:acc_LA}
\acc(\mathrm{o\rightarrow n})&=&\min\left(1,\
\frac{\W{}{\Nst}}{\W{}{\Ost}}\right)\ ,
\end{eqnarray} 
where the Rosenbluth factors are defined as
\begin{eqnarray}
\label{eqn:globalW}
\W{}{\Nst}& =&\frac{\prod_{i=1}^{N}\mathrm{w}_i^\Nst}{\prod_{j=2}^N\mathrm{w}_j^\Nst(1)}
		 \nonumber \\
\W{}{\Ost}&=
		 &\frac{\prod_{i=1}^{N}\mathrm{w}_i^\Ost}{\prod_{j=2}^N\mathrm{w}_j^\Ost(1)}\ .
\end{eqnarray}
Note that the denominators of eq.~(\ref{eqn:globalW}) come from the
bias introduced by eq.~(\ref{eqn:pick_LA}).
In Appendix~\ref{AppB} we prove that the LA method satisfies detailed balance.

%%%%%%%%%%%%%%%%%%%%%%%%%%%%%%%%%%%%%%%%%%%%%%%%%%%%%%%%%%%%%%%%%%%%
\section{Results}
\label{sec-results}
\subsection{Backbone}
%rebridging
%delta phi max
%random picking, NJ, WJ, WJO, WJM
%using delta phi/CPU time as a metric
%Pcross as a metric
%times
We first apply the rebridging scheme to the cyclic peptide
$\mathrm{CG_6C}$\put(-5,15){\line(-1,0){23}}
\put(-5,10){\line(0,1){5}}
\put(-28,10){\line(0,1){5}}
.  Simulation results for the five different
variations of the rebridging scheme were generated.  All simulations
were performed on a Silicon Graphic Indigo$^2$ 195 MHz R10000
workstation.  The system was equilibrated at 298 K. We used an optimized
value of $\Delta\ph{\max}=10^\circ$ in all simulations except
for WJM, in which the optimal value was $\Delta\ph{\max}=30^\circ$. A
probability of 0.05 was assigned for equilibration of either of the two
side chains, NH$_2$ and COOH.  These two side short side chains are
well equilibrated by the method without look-ahead, which is used in
our simulations. We define the acceptance probability $P_\acc$ to be the
ratio of accepted backbone moves
to trial backbone moves.  The efficiency
of the Monte Carlo scheme is measured by the average displacement of
the molecule per CPU time. We define $\Delta\ph{\mathrm{avg}}$ as the
average of the absolute change of torsional angles per trial backbone move:
\begin{eqnarray}
\Delta\ph{\mathrm{avg}}& = & \frac{\sum_{i=1}^{N_\mathrm{trial}}\sum_{j=0}^7|\Delta\ph{j}(i)|}
		    {N_{\mathrm{trial}}}\ .
\end{eqnarray}
This value is a measure of the size of successful moves and the
efficiency of the rebridging scheme.  There is an intrinsic energy
barrier for the $\mathrm{C_{\beta}SSC_{\beta}}$ dihedral angle at
$\ph{\mathrm{C_\beta SSC_\beta}}\simeq 180^\circ$.  The magnitude of
this barrier is estimated to be 5.5-6.5
Kcal$\cdot$mol$^{-1}$~\cite{I_Deem1}.  A barrier-crossing event
happens whenever this angle crosses $\ph{\mathrm{C_\beta
SSC_\beta}}=180^\circ$.  We define the barrier-crossing frequency as
the total number of barrier-crossing events divided by the total
number of backbone moves.  Table~\ref{tab:comp4} lists simulation
results obtained with the five different rebridging methods.

%say something about WJM!
Figure~\ref{fig:CSSC}
shows histograms of the angles \ph{\mathrm{C_\beta SSC_\beta}} observed in these simulations.
%Figure~\ref{fig:CSSCwjk2} shows the histogram observed with WJM and MT.
%The number of multiple rotations $k$ is 2. 
The NJ method yields a left peak that is slightly higher than those from other
methods. The MT method yields the lowest left peak. Although the histograms are
similar, they did not converge to a unique distribution within our
chosen simulation time. This is because barrier crossing was not
frequent enough to produce accurate statistics.

%parallel tempering
To increase the sampling efficiency, we performed a parallel tempering
simulation with 4 systems.  The system temperatures were 298 K, 500 K,
1000 K, and 3000 K.  The rebridging moves were performed using the WJ
biasing method.  The probabilities for proposing swapping moves,
backbone moves, and side chain moves were 0.1, 0.45, and 0.45,
respectively. When a swapping move was chosen, two randomly chosen
adjacent systems were proposed to swap configurations. The
probabilities for swapping the two pairs with lower temperatures were
doubled to accelerate de-correlations.  When a backbone move or a side
chain move was proposed, the system was picked with a probability that
updates the two lowest temperature systems twice as frequently. We do 
this because
of the longer correlation times at lower temperatures. The simulation
consisted of 160000 Monte Carlo cycles.  Each cycle proposed four
swapping or updating moves, chosen at random.  
The whole CPU time taken in this run was
48 hours.  The initial 20000 cycles were discarded to avoid
equilibration effects.

%energy histogram
The swapping moves can occur with sufficient probability only
if the energy histograms of adjacent systems overlap. 
Figure~\ref{fig:e_histo} shows that this condition is satisfied for our
choice of temperatures.
Table~\ref{tab:acc_para}
lists the acceptance probabilities of swapping moves
in this simulation.

%histograms of CSSC at 3000K
%histogram of CSSC at 298K
Figure~\ref{fig:csscpara} shows the distribution of the
$\mathrm{C_\beta SSC_\beta}$ angle observed in the simulation.  The
histogram converged to a unique distribution with very little
simulation data.  After 80000 cycles, the observed distribution was
almost indistinguishable from the one observed at 160000 cycles.  With parallel
tempering, we obtain substantially better statistics in less computation
time. If fact, the computation time was two-thirds of
that used in the single temperature simulations in figure~\ref{fig:CSSC}. 
Note that the
histogram at 3000 K is essentially flat, and so at this temperature
the molecule is free to cross the barrier at 
$\ph{\mathrm{C_\beta SSC_\beta}}\simeq 180^\circ$.  Strong steric repulsion between
hydrogen atoms connected to the adjacent $\mathrm{C_\beta}$ atoms still prevents the
molecule from adopting a conformation with $\ph{\mathrm{C_\beta SSC_\beta}}\simeq
0^\circ$, but this does not hinder equilibration.

\subsection{Side Chains}
%semi-look ahead works well
\label{sec-sidechain}

We performed simulations on the cyclic
CNWKRGDC\put(-5,15){\line(-1,0){65}} \put(-5,10){\line(0,1){5}}
\put(-70,10){\line(0,1){5}}~molecule to test various side chain
regrowth methods.
This medically-relevant molecule has long and bulky side chains.
Simulations were done both on a fixed backbone scaffold and on a backbone
equilibrated with rebridging and parallel tempering. 
First, we fixed the backbone and chose
side chains at random to regrow, using the method without look-ahead
and the SLA method.  We tested the dependence of the equilibration on
the number of trial rotations $n_1$. The backbone was fixed throughout
this simulation.  Figure~\ref{fig:old_and_new} shows the energy as a
function of CPU time during the equilibration period.  Starting from a
high energy configuration, the SLA method with $n_1=100$ or $n_1=10$
reaches equilibrium rapidly.  The non-look-ahead method, however, had
difficulty in finding low energy regions. It took the system with $n_1
=10$ more than 50 minutes to reach low energy configurations. The
system with $n_1=100$, however, never reached equilibrium during the
simulation. Although the associated acceptance probabilities are not
small, the use of $n=100$ results in essentially non-ergodic sampling.
We point out that the
non-look-ahead method equilibrates the system faster with
$n_1=1$ than with
$n_1=10$ or $n_1=100$, although non-look-ahead is always slower than SLA. 

Figure~\ref{fig:old_and_new} may prompt the following question: How do we
determine the optimal value for $n_1$? For short side chains, we
expect that a small $n_1$ will work well. For longer side chains, we expect that
a larger value of $n_1$ will help to explore the torsional space. The
optimal value, therefore, will differ for each side chain.

%compare on W, R, K of CNWKRGDC
%look ahead
We next performed parallel tempering simulations with five systems,
using the SLA, LA, and LARC methods for side chain regrowth. The
backbone moves were performed by the WJ biasing method. The system
temperatures were 298 K, 450 K, 780 K, 1700 K, and 5000 K. 
The
simulation consisted of 100000 Monte Carlo cycles, except in the cases
of $n_1=1$ and $n_1=30$ for SLA and $n_1\times n_2=20\times 10$ for LA, for
which the number of cycles were 200000, 200000, and 60000, respectively.
Each cycle proposed five swapping or updating moves, chosen at
random. The probabilities for proposing swapping moves, backbone
moves, and side chain moves were 0.1, 0.45, and 0.45,
respectively. When a swapping move was proposed, two adjacent systems
were chosen randomly, with the probability of picking system 1, 2, 3,
and 4 equal to $\frac{3}{7}$, $\frac{2}{7}$, $\frac{1}{7}$, and
$\frac{1}{7}$, respectively. When an updating move, either for
backbones or for side chains, was proposed, we chose system 1, 2, 3,
4, and 5 with the probabilities $\frac{3}{8}$, $\frac{2}{8}$,
$\frac{1}{8}$, $\frac{1}{8}$, and $\frac{1}{8}$, respectively. We
focused on the sampling efficiencies for the tryptophan, lysine, and
arginine residues. The lysine residue has a large number, 5, of rigid
units.  The tryptophan residue has an indole group. The arginine
residue has a guanidine group. Both groups are bulky and tend to have
low acceptance probabilities.  For each side chain, we used the total
torsional displacement per computation time, $\Delta\ph{}$/CPU, as an
index to the efficiency. Both side chain moves and swapping moves
contributed to $\Delta\ph{}$. We define the acceptance probability
$P_\acc$ to be the ratio of successful moves to trial moves in a side
chain. The results are summarized in table~\ref{tab:residues}.

Among the four simulations with SLA, the choice $n_1=10$ yields the
best efficiency for lysine, and $n_1=30$ yields the best efficiency
for tryptophan and arginine. Among the four simulations with LA, the
best efficiency for tryptophan is produced when $n_1\times
n_2=10\times 5$.  Lysine and arginine are equilibrated most
efficiently with $n_1\times n_2=10\times 10$. With LARC the efficiency
for tryptophan is the best when $n_1=5$. The
efficiency for lysine is the best when $n_1=10$. Interestingly,
arginine is so difficult to equilibrate, typically having such a low
acceptance probability, that the efficiency was best with $n_1=15$. In
general, LARC is more efficient than LA. Comparing the results from
various methods, we find that tryptophan is equilibrated most
efficiently by SLA, and lysine and arginine are equilibrated most
efficiently by LARC.

%%%%%%%%%%%%%%%%%%%%%%%%%%%%%%%%%%%%%%%%%%%%%%%%%%%%%%%%%%%%%%%%%%%%
\section{Discussion}
\label{sec-discuss}
%rebridging
%note picking one at random bad
%1 additional degree of freedom re concerted rotation
% cound change more, but  will eventually reach some limit
% must be some optimum

%can see subbasins in histogram--we don't want to analyze what they are
%becaues big barrier plus other degrees of freedom slow
%WJM!
Among the five rebridging methods listed in table~\ref{tab:comp4}, WJO
gives the highest acceptance probability, where $P_\acc$ in WJO is
defined to be the probability of accepting a solution other than the
old one. WJO also produces the highest $\Delta\ph{\mathrm{avg}}$.
% the definition of $P_\acc$ for WJO 
The
distribution generated by WJO is the most smooth among the curves,
which shows that it is efficient in sampling local conformations. However, the
CPU time per move for WJO is slightly higher than that for WJ or NJ,
since there is no early rejection in WJO.  
We performed simulations with WJM using different $\Delta\ph{\max}$
and found the optimal value to be
$\Delta\ph{\max}=30^\circ$. Simulations with $\Delta\ph{\max}<30^\circ$
were dominated by smaller moves that lead to infrequent barrier-crossing.
The computation cost per WJM move is roughly proportional to the
number of trial rotations. It is seen in table~\ref{tab:comp4} that
each WJM move takes more than twice the time of a WJ move. Therefore,
WJM is less efficient than the first three schemes in
table~\ref{tab:comp4}. As expected, MT
yields a fairly low acceptance probability. Taking the CPU cost into
consideration, the efficiency of WJ is close to that of WJO.  The efficiency
of NJ is less than WJ and WJO. The WJM method is less efficient than the previous
three schemes. The MT method is the least efficient.

%barrier crossing increased, but not enough for good statistics
Our rebridging scheme is capable of overcoming energy barriers and
promoting the frequency of barrier crossing.  The fourth column in
table~\ref{tab:comp4} lists the barrier-crossing frequency.  The WJ
method yields
the highest barrier-crossing frequency, and WJO yields the lowest. This
is due to the predominance of local moves in WJO. An
accepted move in WJO can be a move that reconfigures six degrees of
freedom only, which is less likely to lead to a barrier-crossing event.

Barrier-crossing is a rare event in a simulation of the 
$\mathrm{CG_6C}$\put(-5,15){\line(-1,0){23}}
\put(-5,10){\line(0,1){5}}
\put(-28,10){\line(0,1){5}}~
peptide. According to the potential of mean force determined by
umbrella sampling, the potential at
$\ph{\mathrm{C_\beta SSC_\beta}}=90^\circ$ is less than that at
$\ph{\mathrm{C_\beta SSC_\beta}}=270^\circ$ by roughly 1
Kcal$\cdot$mol$\mathrm{^{-1}}$~\cite{I_Deem1}. 
We, therefore, expect the left peak to
be substantially higher than the right one.  Our results with
biased rebridging moves are consistent with the potential of mean
force, but the statistics are not good enough.  Because steric
repulsions are severe in our system, the correlation time for other
degrees of freedom is also long, and these degrees of freedom
also slow down the barrier-crossing. We suspect that there is a set of
low energy conformations, separated by low-energy barriers, near
$\ph{\mathrm{C_\beta SSC_\beta}}=270^\circ$.

We have found that parallel tempering is an efficient and automatic
means to overcome these barriers.
The overlap of energy histograms guaranteed reasonable acceptance
probabilities of the swapping moves. These swapping moves transfer
configurations encountered at high temperatures to systems with low
temperatures, thereby helping the low-temperature systems to escape from local energy minima.
Such escape from local minima is important for efficient sampling,
especially in glassy systems with high energy barriers.
%    works very well
%    parallel tempering finds local minima well
%    beacause system is glassy
%      (Hannsman compared on a linear peptide, which has NO big barriers)
Cyclic peptides fall in this category, because of the torsional barriers
and steric repulsions associated with the cyclic constraint. Our results
provide additional evidence that parallel tempering is a powerful tool
for studying glassy systems. Linear peptides, on the other hand, have
a fairly simple free energy landscape, and so they do not benefit
substantially from the parallel tempering approach~\cite{Hansmann}.
%barrier crossing problem solved

%semi-look ahead
For equilibration of side chains, 
we tested whether the inclusion of
torsional interaction energy in the internal
potential is effective. We find that the acceptance probability is lower
and the simulation time is increased through the use of internal biasing.
%U_int doesn't work well
%      torsion is small compared to everything else
%      actually biasing the sampling of wrong regions
Presumably this is because $\mathrm{U^{int}}$ is only a small fraction of the
total interaction energy, and so biasing the torsional angles according to
this term does not
lead to better sampling.
%    old methods not equilibrated on long side chains with bulky groups

CBMC without any look-ahead does not equilibrate long or bulky side
chains as well as does CBMC with look-ahead. The key difference is that
without look-ahead, the head atoms of succeeding units are not
included. Without look-ahead, a chosen rotation, though probably a low
energy configuration for the local atoms, may implicitly put adjacent head atoms in
high energy positions and thereby fail to find the lowest energy region.  Using
fewer twigs in the non-look-ahead method resulted in better equilibration,
as shown by the results for $n_1=1$ in figure~\ref{fig:old_and_new}.
This occurs because with $n_1=1$ the regrowing units have a better chance to miss the
incorrectly identified low energy regions.

%    note different optimal k for different side chains
Increasing the number of twigs raises the acceptance probability in
SLA, LA, and LARC, but at an increased CPU cost. The optimal $n_1$ is
attained when these competing effects are balanced.  We know that for
rougher energy landscapes more trial rotations need to be generated.
From the first row of table \ref{tab:residues}, all three residues
were poorly equilibrated by SLA with $n_1=1$. The torsional
displacement $\Delta\ph{}$ in this case comes mainly from the swapping
moves. Equilibration is improved by using a greater $n_1$, which
increases the acceptance probabilities significantly. However, the
acceptance probability for arginine with $n_1=100$ is lower than that
with $n_1=30$. This means that improving the local sampling does not
always lead to better global sampling, and this in turns implies the
necessity of more significant look-ahead sampling.  The arginine
residue is both long, with four rigid units, and big, with a
guanidine group at the end. Therefore, look-ahead is crucial to bypass
high energy regions.  Comparing the results for SLA with $n_1=10$, LA
with $n_1\times n_2=10\times 10$, and LARC with $n_1=10$, we see both
LA and LARC enhance the acceptance probabilities. The only exception
is the shortest residue, tryptophan, for which LA yields a acceptance
probability slightly lower than that from SLA. Clearly, LARC is
superior to LA, because LARC costs less computation time while
yielding higher acceptance probabilities. For the long residues,
lysine and arginine, LARC yields the highest efficiencies among these
three methods. The results suggest that, for long and bulky side
chains, significant look-ahead is necessary. 
It is not necessary to
use the same regrowth method for all side
chains. 
Indeed, the optimal approach is to use a different regrowth
method for side chains of different identity.  
For short side chains, SLA appears to be optimal.
For longer side chains, LARC is the best method to use.
We believe there may be some cases in which look-ahead is
the only efficient approach for equilibration.
Likely cases are those where there is substantial crowding and 
steric overlap, such as  docking of a drug or signaling molecule to 
a protein receptor site or
binding of antigen by the 
hypervariable region of antibodies.

%%%%%%%%%%%%%%%%%%%%%%%%%%%%%%%%%%%%%%%%%%%%%%%%%%%%%%%%%%%%%%%%%%%%
\section{Conclusion}
\label{sec-conclude}
Peptide function comes primarily from the chemical
functionality of the side chains atoms, although the
side chains themselves are positioned by the
backbone atoms.  For cyclic peptides,
both backbone and side chain atoms are difficult
to equilibrate with standard simulation techniques.
We have described a new and efficient Monte Carlo simulation 
method for complex cyclic peptides.  The combination of
biased, look-ahead Monte Carlo and parallel tempering
leads to rapid and accurate sampling of the relevant
room- or body-temperature conformations.
Specifically, the look-ahead biasing is helpful for equilibrating 
long or bulky side
chains, and the parallel tempering is essential
for crossing torsional-angle free-energy barriers
at a rapid rate.  A variety of details, such as prescreening,
improved Jacobian biasing, semi-look-ahead, and look-ahead, are important
components of the method.

We believe that parallel tempering will prove to be a generally useful
method for simulation of `glassy' atomic systems with multiple,
important conformations separated by large and unpredictable
free energy barriers.  Explicit atom models, which are more
accurate but which also increase the ruggedness of the potential
energy landscape, are naturally treated within this approach.
We expect that application of our peptide simulation method to high-density
or crowded situations, such as peptide-receptor or antibody-antigen binding events,
will further demonstrate the efficiency and power of our approach.

\section*{Acknowledgments}
We thank Marco Falcioni for many useful discussions.
This research was supported by the National Science Foundation through
grants CTS--9702403 and CHE-9705165.

%\newpage
\appendix
\section{The Jacobian in the Rebridging Scheme}
\label{AppA}

In the rebridging scheme, each solution should be weighted 
by a Jacobian to correct for the non-uniform distribution of the
angles $\ph{1},\ldots,\ph{6}$ generated by the non-linear solution of  the
geometrical problem. We derive the Jacobian here from the
classical partition function. 

We initially consider a simple cyclic molecule with only N backbone atoms and N backbone
torsional degrees of freedom. 
This assumption is relaxed at the end to accommodate the complicated
backbone and side chain geometry of a real peptide.
The momentum part of the partition function can be integrated out if
we assume that the bond length and angle constraints are enforced by
springs with infinite force constants.
We, thus, focus on the configurational part.
The configurational partition function is
\begin{eqnarray}
\label{eqn:fixend}
Z &\equiv & \int \drs{N}\boltz{\mathrm{U}} \nonumber \\
	& = &\int \drs{N}\drv{N+1}\drv{N+2}\drv{N+3}
	\delta^{3}(\rv{N+1}-\rv{1})\delta^{3}(\rv{N+2}-\rv{2})
		\delta^{3}(\rv{N+3}-\rv{3})\boltz{\mathrm{U}}\; ,
	\nonumber \\
\end{eqnarray}
where we have introduced three vector delta functions to
account for the cyclic constraint.
The choice of fixed-end constraints is not unique. We will discuss an alternative form later in this section.

We start the derivation by performing a transformation from $\rv{}^N$ to $\yv{}^N$
\begin{eqnarray}
\yv{1} & = & \rv{1} \nonumber \\
\yv{i} & = & \rv{i} - \rv{i-1} \mbox{, }i=2,\ldots, N+3\ .
\end{eqnarray}
The Jacobian of this transformation is unity.
We transform again from $\yv{}^N$ to local coordinates.
We define $l_i=|\yv{i}|$ and \thet{i} to be the angle formed by \yv{i} and \yv{i+1}.
We transform from \yv{2} to $l_2$ and \uv{2}, where
\begin{eqnarray}
\uv{2} &\equiv          & \frac{\yv{2}}{|\yv{2}|}\ . \nonumber \\
\end{eqnarray}
Then we transform from \yv{3} to $l_3,\ \thet{2},$ and $\gamma_2$, where
$\gamma_2$ is the azimuthal angle of $\hat {\bf u}_3$ in a spherical
coordinate system defined with $\hat {\bf u}_2$ as the $z$-axis.  The
angle is measured with respect to the plane defined by $\hat {\bf
u}_2$ and $\hat {\bf e}_3$, the fixed laboratory $z$-axis.  
We further transform \yv{i}
to a spherical coordinate system $l_i,\ \thet{i-1},$ and \ph{i-1}, $i=4,\ldots,N+3$.
With this transformation, we obtain
\begin{eqnarray}
\label{eqn:car_to_int}
Z &=    &\int
\dyv{1}{l_2}^2 dl_2 d\uv{2}{l_3}^2 dl_3 d\thet{2}\sin\thet{2}d\gamma_2\int
\prod_{i=3}^{N+2}dl_{i+1} d\thet{i} d\ph{i}
\nonumber\\
& \times&
\delta^3\left(\sum_{j=2}^{N+1}\yv{j}\right)\,
\delta^3(\yv{N+2}-\yv{2}) \, \delta^{3}(\yv{N+3}-\yv{3})
\nonumber\\
& \times &
\mathrm{J} \left(\frac{\yv{4},\ldots,\yv{N+3}}{l_4,\ldots, l_{N+3},\
\thet{3},\ldots,\thet{N+2},\ \ph{3},\ldots,\ph{N+2}}\right)
\boltz{\mathrm{U}}\; .
\end{eqnarray}

The Jacobian is simply
$\mathrm{J}=\prod_{i=3}^{N+2} {l_{i+1}}^2 \sin\thet{i}$. The fast coordinates $l_i$
and \thet{i} are fixed due to the strong harmonic potentials. We
denote the equilibrium values of $l_i$ and \thet{i} by $l_i^0$ and
$\thet{i}^0$, respectively.  With very large spring constants, the
dependence of the integrand along these coordinates can effectively be
replaced with delta functions.  Therefore
\begin{eqnarray}
Z & = &\mathrm{C'''}\int
\dyv{1}{l_2}^2dl_2d\uv{2}
{l_3}^2dl_3d\thet{2}\sin\thet{2}d\gamma_2\ \delta(l_2-l_2^0)
\delta(l_3-l_3^0)\delta(\thet{2}-\thet{2}^0) \nonumber \\
&\times & 
\int \prod_{i=3}^{N+2}dl_{i+1} d\thet{i}
d\ph{i}\ \mathrm{J}(l_4,\ldots, l_{N+3},\
\thet{3},\ldots,\thet{N+2}) \boltz{\mathrm{U}_0} \nonumber \\
&\times & \delta^3\left(\sum_{j=2}^{N+1}\yv{j}\right)
\delta^3(\yv{N+2}-\yv{2})\delta^3(\yv{N+3}-\yv{3})\nonumber\\
&\times &
\prod_{k=4}^{N}\delta(l_k-l_k^0)\prod_{k'=3}^{N-1}\delta(\thet{k'}-\thet{k'}^0)
\delta(|\sum_{k''=2}^{N}\yv{k''}|-l_1^0) \nonumber  \\
&\times  &
\delta\left(\cos^{-1}\frac{\yv{2}\cdot(-\sum_{i'=2}^{N}\yv{i'})}
	{|\yv{2}||\sum_{i''=2}^{N}\yv{i''}|}-\thet{1}^0\right)
\delta\left(\cos^{-1}\frac{\yv{N}\cdot(-\sum_{j'=2}^{N}\yv{j'})}
	{|\yv{N}||\sum_{j''=2}^{N}\yv{j''}|}-\thet{N}^0\right)\; ,
\end{eqnarray}
where $\mathrm{U}_0$ is the potential energy measured at the ground
configuration of these hard coordinates. We now integrate over the hard
coordinates $l_i$ and \thet{i}.  The Jacobian is simply a constant and can be
taken out of the integral.  Since the constraint
$\delta^3(\sum_{i=2}^{N+1}\yv{i})$ holds, we can replace every
$-\sum_{i=2}^N\yv{i}$ with \yv{N+1}. Similarly, we can replace \yv{2}
with \yv{N+2}. Replacing the arguments in the last two delta functions
with \thet{N+1} and \thet{N}, respectively, we obtain
\begin{eqnarray}
Z & = &\mathrm{C''}\int
\dyv{1}d\uv{2}d\gamma_2
\int\prod_{i=N}^{N+2}dl_{i+1} d\thet{i} d\ph{i}\  
\boltz{\mathrm{U}_0}
 \nonumber \\
&\times & 
\delta^3\left(\sum_{j=2}^{N+1}\yv{j}\right)
\delta^3(\yv{N+2}-\yv{2})\delta^3(\yv{N+3}-\yv{3})
\delta(|\yv{N+1}|-l_1^0) \nonumber \\
&\times &
\delta\left(\thet{N+1}-\thet{1}^0\right)
\delta\left(\thet{N}-\thet{N}^0\right)\; .
\end{eqnarray}
We use the equalities
\begin{eqnarray}
\delta^3(\yv{N+2}-\yv{2}) &=
&\delta(l_{N+2}-l_2)\delta^2(\uv{N+2}-\uv{2})/{l_{N+2}}^2 \nonumber \\
\delta^3(\yv{N+3}-\yv{3}) &=
&\delta(l_{N+3}-l_3)\delta^2(\uv{N+3}-\uv{3})/{l_{N+3}}^2
\nonumber
\end{eqnarray}
to integrate over $l_{N+1}$, $l_{N+2}$, $l_{N+3}$, \thet{N}, and
\thet{N+1} to obtain
\begin{eqnarray}
Z & = &\mathrm{C'}
\int\dyv{1}d\uv{2}d\gamma_2\int\prod_{i=3}^{N+2}d\ph{i}\int d\thet{N+2}
\;\boltz{\mathrm{U}_0}
\nonumber \\
& \times & 
\delta^3\left(\sum_{j=2}^{N+1}\yv{j}\right)
\delta^2(\uv{N+2}-\uv{2})\delta^2(\uv{N+3}-\uv{3})\ .
\end{eqnarray}
Note that
\begin{eqnarray}
\label{eqn:phi2_thet2}
\int d\thet{N+2}\delta^2(\uv{N+3}-\uv{3}) & = &
\int d\thet{N+2}\delta(\gamma_{N+2}-{\gamma_2})
	\delta(\thet{N+2}-\thet{2})/\sin\thet{2} \nonumber \\
& =     & \delta(\gamma_{N+2}-{\gamma_2})/\sin\thet{2}
\; .
\end{eqnarray}
where 
\begin{eqnarray}
\thet{2}& = & \left|\cos^{-1}\frac{\yv{3}\cdot\yv{N+2}}
	{|\yv{3}||\yv{N+2}|}\right|\ ,
\end{eqnarray}
and $\gamma_{N+2}$ and $\gamma_2$ are the azimuthal angles of
$\hat {\bf u}_{N+3}$ and $\hat {\bf u}_3$ in a spherical coordinate
system defined with $\hat {\bf u}_{N+2}=\uv{2}$ as the $z$-axis.  The angles
are measured with respect to the plane defined by $\hat {\bf u}_2$
and $\hat {\bf e}_3$.
Integrating over \thet{N+2}, we obtain
\begin{eqnarray}
\label{eqn:cyclic}
Z & =
 \mathrm{C}&\int\dyv{1}d\uv{2}d\gamma_2\int\prod_{i=3}^{N+2}d\ph{i}
 \nonumber \\
&\times & \boltz{\mathrm{U}_{0}}
\delta^3\left(\sum_{j=2}^{N+1}\yv{j}\right)
\delta^2(\uv{N+2}-\uv{2})\delta(\gamma_{N+2}-\gamma_2)\; .
\end{eqnarray}
This is the partition function of a classical, cyclic molecule. We see
that it is an integral over torsional space with delta function constraints. These
constraints cause an intrinsically non-uniform
distribution of every torsional angle, even in the absence of any
energy of interaction.  It is convenient to transform the last six
torsional coordinates to the variables \rv{N+1}, \uv{N+2}, and
$\ph{N+2}$ and to integrate over these six coordinates. Then
\begin{eqnarray}
\label{eqn:6deltapar}
Z & = & \mathrm{C}\int\dyv{1}d\uv{2}d\gamma_2
\int\prod_{i=3}^{N-4}d\ph{i}\int d\rv{N+1}d\uv{N+2}d\gamma_{N+2}
\nonumber \\
& \times &\sum_{k=1}^{k_\mathrm{s}}
\left\{\mathrm{J}_k\left(\frac{\ph{N-3},\ldots,\ph{N+2}}
	{\rv{N+1},\uv{N+2},\gamma_{N+2}}\right)\boltzm{\mathrm{U}_0(k)}\right\}
\delta^3\left(\sum_{j=2}^{N+1}\yv{j}\right)
\delta^2(\uv{N+2}-\uv{2})\delta(\gamma_{N+2}-\gamma_2) \nonumber
\\
& = & \mathrm{C}\int\dyv{1}d\uv{2}d\gamma_2
\int\prod_{i=3}^{N-4}d\ph{i}\
\sum_{k=1}^{k_\mathrm{s}}
\left\{\mathrm{J}_k\left.\left(\frac{\ph{N-3}\ldots\ph{N+2}}{\rv{N+1},\uv{N+2},\gamma_{N+2}}
		\right)\right|_{\rv{1},\uv{2},\gamma_2}
\boltzm{\mathrm{U}_0(k)}\right\}\ .
\end{eqnarray}
The index $k$ labels the solutions $\{\ph{N-3},\ldots,\ph{N+2}\}$ that
satisfy the fixed-end constraints.  The summation accounts for the
fact that multiple solutions are possible. 
In the rebridging scheme, we always relabel $\ph{N-3},\ldots,\ph{N+2}$ as
$\ph{1},\ldots,\ph{6}$ and $\rv{1},\uv{2},\mbox{ and }\gamma_2$
as $\rv{5},\uv{6},\mbox{ and }\gamma_6$.  From 
eq.~(\ref{eqn:6deltapar}) it is clear that each solution must be given a weight, which
is the Jacobian. 

The $6\times6$ Jacobian is actually the determinant of a $5\times5$
matrix, since the last torsional angle does not affect \rv{5} or
\uv{6}. Therefore,
\[\frac{\partial \rv{5}}{\partial \ph{6}}
= \frac{\partial
\uv{6}}{\partial \ph{6}}=0\ .\] 
We also know that
\[\frac{\partial \gamma_6}{\partial \ph{6}}=1\ .\]
So we obtain
\begin{eqnarray}
\label{eqn:jac_RBsim}
\mathrm{J}\left(\frac{\ph{1},\ph{2},\ph{3},\ph{4},\ph{5},\ph{6}}
	{\rv{5},\ \uv{6},\ \gamma_6}\right)
& =     & \frac{\uv{6}\cdot\hat{\mathbf{e}}_{3}}{\det|B|}\nonumber \\
B_{ij} & = & [\uv{j}\times(\rv{5}-\rv{j})]_i \mbox{, if } j\leq3 
							\nonumber \\
  & = & [\uv{j}\times\uv{6}]_{j-3} \mbox{, if } j=4\mbox{ or }5.
\end{eqnarray}

Since the Jacobian is independent of \ph{6}, we might conjecture that
it is also independent of \ph{1}.  The reason is that the Jacobian
should not depend on the direction that we choose for the labeling of
the rigid units.  Hoffmann and Knapp derived a
$4\times 4$ Jacobian depending only on \ph{2}, \ph{3}, \ph{4}, and
\ph{5} for case 6 of table~\ref{tab:0pro}~\cite{Hoffmann}. 
We will show that, with
suitable choice of end-constraint variables, a $4\times 4$ matrix can
be derived in all cases. The idea is to choose a set of end
coordinates that are almost independent of \ph{1}.

Integrating eq.~(\ref{eqn:cyclic}) over \ph{N+2}, we obtain
\begin{eqnarray}
\label{eqn:cyclic5}
Z & = \mathrm{C}&\int\dyv{1}d\uv{2}d\gamma_2\int\prod_{i=3}^{N+1}d\ph{i}\;
\boltz{\mathrm{U}_0} \delta^3\left(\sum_{j=2}^{N+1}\yv{j}\right)
\delta^2(\uv{N+2}-\uv{2})\; .
\end{eqnarray}
Let $\Delta\rv{}= \rv{N+1} - \rv{N-3}$ and introduce the following end
coordinates
\begin{eqnarray}
R       & = &|\Delta\rv{}| \nonumber \\
\thet{ b}& = & 
	\left|\cos^{-1}\left(\frac{\Delta\rv{}\cdot\uv{N-3}}{R}\right)\right|
		\nonumber \\
\ph{ b}         & = & \mbox{the torsional angle of $\Delta$\rv{} in local
		coordinates of unit $N-3$} \nonumber \\
\thet{\mathrm{e}} & = & 
\left|\cos^{-1}\left(\frac{\Delta\rv{}\cdot\uv{N+2}}{R}\right)\right|
	 \nonumber \\
\ph{\mathrm{e}} & = & 
	\mbox{the torsional angle defined by \uv{N-3}, $\Delta$\rv{},
		and \uv{N+2}}\ .
\end{eqnarray}
Note that $R$, \thet{ b}, \thet{\mathrm{e}}, 
and \ph{\mathrm{e}} are independent of
\ph{N-3} and that \ph{ b} is linear in \ph{N-3}.  Substituting these
coordinates into eq.~(\ref{eqn:cyclic5}), we obtain
\begin{eqnarray}
\label{eqn:cyclic5A}
Z & =
\mathrm{C}&\int\dyv{1}d\uv{2}d\gamma_2\int\prod_{i=3}^{N+1}d\ph{i}\boltz{\mathrm{U}_0}
\frac{1}{R^2\sin\thet{ b}\sin\thet{\mathrm{e}}}
 \nonumber \\ 
&\times&
\delta(R - |\rv{1} - \rv{N-3}|)\delta(\ph{ b}-\ph{ b}^0)
	\delta(\thet{ b}-\thet{ b}^0)
	\delta(\thet{\mathrm{e}}-\thet{\mathrm{e}}^0)
	\delta(\ph{\mathrm{e}}-\ph{\mathrm{e}}^0)\; .
\end{eqnarray}
Here 
\begin{eqnarray}
\thet{ b}^0     & = & \left|\cos^{-1}\left(
		\frac{(\rv{1} - \rv{N-3})\cdot\uv{N-3}}{|\rv{1} - \rv{N-3}|}\right)\right|
		\nonumber \\
\ph{ b}^0       & = & \mbox{the torsional angle of $\rv{1}-\rv{N-3}$ in local
		coordinates of unit $N-3$} \nonumber \\
\thet{\mathrm{e}}^0 & = &  \left|\cos^{-1}\left(\frac{(\rv{1} -
		\rv{N-3})\cdot\uv{2}}{|\rv{1} - \rv{N-3}|}\right)\right|
			\nonumber \\
\ph{\mathrm{e}}^0  & = & \mbox{the torsional angle defined by \uv{N-3}, $\Delta$\rv{},
		and \uv{2}} \ .
\end{eqnarray}
Transforming coordinates from $\ph{N-3},\ldots,\ph{N+1}$ to $R$,
\thet{ b}, \ph{ b}, $\thet{\mathrm{e}}$, and
$\ph{\mathrm{e}}$, we obtain
\begin{eqnarray}
\label{eqn:5by5jac}
Z & = & \mathrm{C}\int\dyv{1}d\uv{2}d\gamma_2
\int\prod_{i=3}^{N-4}d\ph{i}
\int dRd\thet{ b}d\ph{ b}
d\thet{\mathrm{e}}d\ph{\mathrm{e}}\ \frac{1}{R^2\sin\thet{ b}\sin\thet{\mathrm{e}}}
	\nonumber \\
& \times &
	\delta(R - |\rv{1} - \rv{N-3}|)
	\delta(\thet{ b}-\thet{ b}^0)\delta(\ph{ b}-\ph{ b}^0)
	\delta(\thet{\mathrm{e}}-\thet{\mathrm{e}}^0)\delta(\ph{\mathrm{e}}-\ph{\mathrm{e}}^0)
								\nonumber \\
	&\times &
\sum_{k=1}^{k_\mathrm{s}}
\left\{\mathrm{J}_k\left.\left(\frac{\ph{N-3},\ldots,\ph{N+1}}
	{R,\ \thet{ b},\ \ph{ b},\ \thet{\mathrm{e}},\ \ph{\mathrm{e}}}
	\right)\right|_{|\rv{1} - \rv{N-3}|,\thet{ b}^0,\ph{ b}^0,
	\thet{ b}^0,\ph{\mathrm{e}}^0}\boltzm{\mathrm{U}_0(k)}\right\} 
								 \nonumber \\
& = & \mathrm{C}\int\dyv{1}d\uv{2}d\gamma_2
\int\prod_{i=3}^{N-4}d\ph{i}\; 
	\frac{1}{R^2\sin\thet{ b}\sin\thet{\mathrm{e}}}\nonumber \\
& \times &
\sum_{k=1}^{k_\mathrm{s}}
\left\{\mathrm{J}_k\left.\left(\frac{\ph{N-3},\ldots,\ph{N+2}}
	{R,\ \thet{ b},\ \ph{ b},\ \thet{\mathrm{e}},\ \ph{\mathrm{e}}}
	\right)\right|_{|\rv{1} - \rv{N-3}|,\thet{ b}^0,\ph{ b}^0,\thet{\mathrm{e}}^0,\ph{\mathrm{e}}^0}
\boltzm{\mathrm{U}_0(k)}\right\}\; .
\end{eqnarray}
The Jacobian can be rewritten as
\begin{equation}
\mathrm{J}      =  \frac{1}{R^2\sin\thet{ b}\sin\thet{\mathrm{e}}
			|\det(\mathrm{B''})|}\; ,
\end{equation}
where
\begin{eqnarray}
\label{eqn:jacb''}
\mathrm{B''}_{1j}& =&  \frac{\partial R}{\partial \ph{N-4+j}} \, ,
\mathrm{B''}_{2j} = \frac{\partial \thet{ b}}{\partial \ph{N-4+j}} \, ,
\mathrm{B''}_{3j} = \frac{\partial \ph{ b}}{\partial \ph{N-4+j}}\, , \nonumber \\
\mathrm{B''}_{4j}& =& \frac{\partial \thet{\mathrm{e}}}{\partial \ph{N-4+j}} \, ,
 \mathrm{B''}_{5j} = \frac{\partial \ph{\mathrm{e}}}{\partial \ph{N-4+j}} 
 \;\;\mathrm{, } \;\;j = 1,\ldots, 5\; .
\end{eqnarray}
The first column of $\mathrm{B''}$ has only one non-zero element, which is
$\mathrm{B''}_{31}=1$.  Taking the cofactor of $\mathrm{B'}_{21}$, 
the determinant can be
replaced with that of a $4\times 4$ matrix
\begin{eqnarray}
\label{eqn:jacb'}
\mathrm{B'}_{1j}& =&  \frac{\partial R}{\partial \ph{N-3+j}} \, ,
\mathrm{B'}_{2j} = \frac{\partial \thet{ b}}{\partial \ph{N-3+j}} \, ,
\mathrm{B'}_{3j} = \frac{\partial \thet{\mathrm{e}}}{\partial \ph{N-3+j}} \, ,
\nonumber \\ 
 \mathrm{B'}_{4j} &= &\frac{\partial \ph{\mathrm{e}}}{\partial \ph{N-3+j}} 
 \;\;\mathrm{, } \;\;j = 1,\ldots, 4\; . 
\end{eqnarray}

It is easy to extend our approach to include side chains and
constrained torsional angles.  Following an approach parallel to
eqs.~(\ref{eqn:fixend})--(\ref{eqn:car_to_int}), we obtain an integral
with additional degrees of freedom contributed by side chains. These
degrees of freedom are not constrained, and they can be integrated out
first.  Therefore, we can simply replace $\mathrm{U}$ with an
influence functional. The final form of
the Jacobian is unaffected. For peptides, rotation about the C-N bond in the
amide group is governed by a large force constant.  In our simulation, we
constrain these torsional degrees of freedom as well.
Each constrained bond adds a delta function to eq. (\ref{eqn:cyclic}).
Let $A$ be the set of \ph{i} that are constrained. Then
\begin{eqnarray}
\label{eqn:cyclic_pi}
Z & = \mathrm{C}&\int\dyv{1}d\uv{2}d\gamma_2\int\prod_{i=3}^{N+2}d\ph{i}\ \boltz{\mathrm{U}_0}
\delta^3\left(\sum_{j=2}^{N+1}\yv{j}\right)
\delta^2(\uv{N+2}-\uv{2})\delta(\gamma_{N+2}-\gamma_2)
\nonumber \\
& \times&\prod_{k\in A}\delta(\ph{k}-\ph{k}^0)\; .
\end{eqnarray}

Let $G(l,N+2)$  denote the last $l$ flexible torsional angles from \ph{3} to
\ph{N+2}. Integrating out the other degrees of freedom, we obtain
\begin{eqnarray}
\label{eqn:6deltapar_pi}
Z & = & \mathrm{C}\int\dyv{1}d\uv{2}d\gamma_2
\int\left(\prod_{i\notin \left[A\cup G(6,N+2)\right]}d\ph{i}\right)
\nonumber \\
&\times &
\sum_{k=1}^{k_{s}}
\left\{\mathrm{J}_k\left.\left(\frac{G(6,N+2)}{\rv{N+1},\uv{N+2},\gamma_{N+2}}
		\right)\right|_{\rv{1},\uv{2},\gamma_2}
		\boltzm{\mathrm{U}_0(k)}\right\}\; .
\end{eqnarray}

If \ph{N+2} is constrained, \rv{N}, \rv{N+1}, \rv{N+2}, and \rv{N+3}
define a rigid unit.  The corresponding fixed-end coordinates in our
algorithm are chosen to be \rv{N}, \uv{N+1}, and $\gamma_{N+1}$, instead
of \rv{N+1}, \uv{N+2}, and $\gamma_{N+2}$. This apparent difference
causes no ambiguity, since both sets define the same rigid unit.  The
Jacobian between these two sets is unity.

Relabeling the torsional angles in $G(6,\ N+2)$ by
$\ph{1},\ldots,\ph{6}$, we recover the Jacobian in eq.~(\ref{eqn:jac_RB}).
The $4\times4$ Jacobian in this case can be derived analogously. The
final result, which is numerically equal to eq.~(\ref{eqn:jac_RB}), is
\begin{eqnarray}
\label{eqn:jac_4x4}
\mathrm{J}
& =     & \frac{1}{R^2\sin\thet{ b}\sin\thet{\mathrm{e}}
			|\det(\mathrm{B'})|}\; . \nonumber 
\end{eqnarray}
The components of $\mathrm{B'}$ are given below:
\begin{eqnarray}
\mathrm{B'}_{1j} & = &\frac{\partial R}{\partial \ph{j}}
		 =  \frac{1}{R}\frac{\partial\Delta\rv{}}{\partial \ph{j}}\cdot\Delta\rv{}
	\nonumber \\
\mathrm{B'}_{2j} & = &\frac{\partial\thet{ b}}{\partial\ph{j}}
		=\frac{-1}{R\sin\thet{ b}}
	\left[-\frac{1}{R}\mathrm{B'}_{1j}\Delta\rv{}\cdot\uv{1}+
		\frac{\partial \Delta\rv{}}{\partial
	\ph{j}}\cdot\uv{1}\right] \nonumber \\
\mathrm{B'}_{3j} & = &\frac{\partial\thet{\mathrm{e}}}{\partial\ph{j}}
=       \frac{-1}{R\sin\thet{\mathrm{e}}}\left[\frac{-1}{R}\mathrm{B'}_{1j}\Delta\rv{}\cdot\uv{6}
		+\frac{\partial\Delta\rv{}}{\partial\ph{j}}\cdot\uv{6}
		+\Delta\rv{}\cdot(\uv{j}\times\uv{6})\right] \nonumber \\
\mathrm{B'}_{4j} & = & \frac{\partial\ph{ b}}{\partial\ph{j}}
	\nonumber \\
&=      &\frac{-1}{R^2\sin\ph{\mathrm{e}}\sin\thet{ b}\sin\thet{\mathrm{e}}}
	\left\{\frac{(\uv{1}\times\Delta\rv{})\cdot(\Delta\rv{}\times\uv{6})}
	{R^2\sin\ph{\mathrm{e}}\sin\thet{ b}\sin\thet{\mathrm{e}}}\right.
\nonumber \\
& &     \times\left(2\frac{\partial \Delta\rv{}}{\partial \ph{j}}\cdot\Delta\rv{}
	\sin\thet{ b}\sin\thet{\mathrm{e}}
	+R^2\cos\thet{ b}\sin\thet{\mathrm{e}}\mathrm{B'}_{2j}
	+R^2\sin\thet{ b}\cos\thet{\mathrm{e}}\mathrm{B'}_{3j}\right)
\nonumber \\
&       &\left.
	+\left[(\uv{1}\times\frac{\partial \Delta\rv{}}{\partial\ph{j}})
				\cdot(\Delta\rv{}\times\uv{6})
	+(\uv{1}\times\Delta\rv{})
	\cdot\left(\frac{\partial\Delta\rv{}}{\partial \ph{j}}\times\uv{6}
		+\Delta\rv{}\times(\uv{j}\times\uv{6})\right)\right]
	\right\}\; , \nonumber \\
\end{eqnarray}
where 
\begin{eqnarray}
\frac{\partial\Delta\rv{}}{\partial \ph{j}}&=&
\uv{j}\times(\rv{5\mathrm{t}}-\rv{j\mathrm{h}})\; .
\nonumber
\end{eqnarray}
The quantities needed to calculate $\mathrm{B'}$ are \uv{i},
$\Delta\rv{}$, \thet{ b}, \thet{\mathrm{e}}, and $\sin\ph{\mathrm{e}}$.

%%%%%%%%%%%%%%%%%%%%%%%%%%%%%%%%%%%%%%%%%%%%%%%%%%%%%%%%%%%%%%%%%%%%%
\section{Detailed Balance for LA} 
\label{AppB}
In this appendix we prove that the LA method satisfies detailed balance. The
proof for LARC can be done analogously and is not presented here.
In our algorithm, the old Rosenbluth factor ${\mathrm W^\Ost}$ is not
evaluated until all units have been given new positions.  In fact,
${\mathrm W^\Ost}$ can be calculated at any time.  In the proof, we
calculate the partial old Rosenbluth factor $\mathrm{w}_i^\Ost$ of unit $i$
once a new proposed move for unit $i$ is made.  We first derive the probability
for proposing a forward move of the first unit.  By analogy, we derive
the probability for proposing a reverse move of the first unit.  Since
we generate both the old Rosenbluth factor and the new Rosenbluth factor in
a random way, their probabilities should be included.  This is the so-called 
super detailed balance condition~\cite{Frenkel}.  We will show
that LA satisfies super detailed balance.

Let $\alpha_{1}\left(\mathrm{o\rightarrow n};\;
\{\ph{1}^\Nst(\alpha)\}, \{\ph{2}^\Nst(\alpha,\ \gamma)\},
\{\ph{1}^\Ost(\alpha')\},\{\ph{2}^\Ost(\alpha',\ \gamma')\}\right)$ be the
probability of proposing a move from $\ph{1}^\Ost(1)$ to
$\ph{1}^\Nst(1)$, given $\{\ph{1}^\Nst(\alpha)\}, \{\ph{2}^\Nst(\alpha,\ \gamma)\},
\{\ph{1}^\Ost(\alpha')\}$, and $\{\ph{2}^\Ost(\alpha',\ \gamma')\}$.  
Consider the following three events:
\begin{enumerate}
\item
Generating $n_1 n_2$ new twigs, which has the probability 
\[\prod_{\alpha=1}^{n_1}p_1^\Nst(\alpha)\prod_{\gamma=1}^{n_2}p_2^\Nst(\alpha,\
\gamma)\ .\]
\item
Picking a new twig, which has the probability $q_1^\Nst(1)$.
\item
Generating $n_1 n_2$ old twigs, which has the probability
is
\[\prod_{\gamma'=1}^{n_2}p_2^\Ost(1,\ \gamma')
	\prod_{\alpha'=2}^{n_1}\left(p_1^\Ost(\alpha')
	\prod_{\gamma'=1}^{n_2}p_2^\Ost(\alpha',\ \gamma')\right)\ .\]
\end{enumerate}
The probability of the whole event,
$\alpha_{1}\left(\mathrm{o\rightarrow n};\; \{\ph{1}^\Nst(\alpha)\}, \{\ph{2}^\Nst(\alpha,\
j)\}, \{\ph{1}^\Ost(\alpha')\},\{\ph{2}^\Ost(\alpha',\
\gamma')\}\right)$, 
is the product of these three probabilities.  Multiplying the three terms
together, we obtain
\begin{eqnarray}
\label{eqn:dbal_on}
\prod_{\alpha=1}^{n_1}\left(p_1^\Nst(\alpha)
	\prod_{\gamma=1}^{n_2}p_2^\Nst(\alpha,\ \gamma)\right)\times
	q_1^\Nst(1)\times\prod_{\gamma'=1}^{n_2}p_2^\Ost(1,\ \gamma')
	\prod_{\alpha'=2}^{n_1}\left(p_1^\Ost(\alpha')
	\prod_{\gamma'=1}^{n_2}p_2^\Ost(\alpha',\ \gamma')\right)\; .
\end{eqnarray}

Similarly, the probability of proposing the reverse move,\newline
$\alpha_1\left(\mathrm{n\rightarrow o};\; \{\ph{1}^\Ost(\alpha')\}, \{\ph{2}^\Ost(\alpha',\
\gamma')\}, \{\ph{1}^\Nst(\alpha)\},\{\ph{2}^\Nst(\alpha,\
 \gamma)\}\right)$,
is
\begin{eqnarray}
\label{eqn:dbal_no}
\prod_{\alpha'=1}^{n_1}\left(p_1^\Ost(\alpha')
	\prod_{\gamma'=1}^{n_2}p_2^\Ost(\alpha',\ \gamma')\right)\times
	q_1^\Ost(1)\times\prod_{\gamma=1}^{n_2}p_2^\Nst(1,\ \gamma)
	\prod_{\alpha=2}^{n_1}\left(p_1^\Nst(\alpha)
	\prod_{\gamma=1}^{n_2}p_2^\Nst(\alpha,\ \gamma)\right)\ .
\end{eqnarray}

We define $\mathrm{{U'}_1^{ext}}$ as the external energy for unit 1 in
the old configuration and $\mathrm{U_1^{ext}}$ as the external
energy for unit 1 in the new configuration.  Taking the ratio of
eq.~(\ref{eqn:dbal_on}) and eq.~(\ref{eqn:dbal_no}),
most of the
probabilities for generating the twigs cancel. Replacing
$q_1^\Nst(1)$ and $q_1^\Ost(1)$ with eq.~(\ref{eqn:pick_LA}), we obtain
\begin{eqnarray}
\label{eqn:dtbalance_1}
\frac{
\alpha_1(\mathrm{o\rightarrow n};\; \{\ph{1}^\Nst(\alpha)\}, \{\ph{2}^\Nst(\alpha,\ \gamma)\}, 
	\{\ph{1}^\Ost(\alpha')\},\{\ph{2}^\Ost(\alpha',\ \gamma')\})}
{\alpha_1(\mathrm{n\rightarrow o};\; \{\ph{1}^\Ost(\alpha')\}, \{\ph{2}^\Ost(\alpha',\ \gamma')\}, 
	\{\ph{1}^\Nst(\alpha)\}, \{\ph{2}^\Nst(\alpha,\ \gamma)\})}
	\nonumber \\
	=\frac{p_1^\Nst(1)}{p_1^\Ost(1)}
	\frac{\boltzm{\mathrm{U_1^{ext,n}}(1)}}{\boltzm{\mathrm{{U}_1^{ext,o}}(1)}}
	\mathrm{\frac{w_2^\Nst(1)}{w_2^\Ost(1)}}
	\mathrm{\frac{w_1^\Ost}{w_1^\Nst}}
	\nonumber \\
	= 
	\frac{\boltz{\mathrm{U}_1^\Nst}}{\boltz{\mathrm{U}_1^\Ost}}
	\mathrm{\frac{w_2^\Nst(1)}{w_2^\Ost(1)}
	\frac{w_1^\Ost}{w_1^\Nst}}\; ,
\end{eqnarray}
where we have used eq.~(\ref{eqn:p_int_1}) to obtain the last line.
Similarly, we can obtain the ratio of probabilities for subsequent units.
The ratio of the transition probabilities is the product of these ratios and
the ratios of the
acceptance probabilities. Multiplying eq.~(\ref{eqn:dtbalance_1}) for
each unit and using eqs.~(\ref{eqn:acc_LA}) and (\ref{eqn:globalW}), 
we find that super detailed balance is satisfied:
\begin{eqnarray}
\frac{\alpha(\mathrm{o\rightarrow n})\acc(\mathrm{o\rightarrow n)}}
	{\alpha(\mathrm{n\rightarrow o})\acc(\mathrm{n\rightarrow o)}}
	&= &\frac{\prod_{i=1}^{N}\boltz{\mathrm{U}_i^\Nst}}
			{\prod_{i'=1}^{N}\boltz{\mathrm{U}_{i'}^\Ost}}\times
		\frac{\prod_{j=2}^N\mathrm{w}_j^\Nst(1)}
			{\prod_{j'=2}^N\mathrm{w}_{j'}^\Ost(1)}
		\frac{\prod_{k=1}^{N}\mathrm{w}_k^\Ost}
		{\prod_{k'=1}^{N}\mathrm{w}_{k'}^\Nst}
		\frac{\mathrm{W}^\Nst}{\mathrm{W}^\Ost} \nonumber \\
	&= & \frac{\boltz{\mathrm{U}^\Nst}}{\boltz{\mathrm{U}^\Ost}}\ .
\end{eqnarray}

\bibliography{rebridge}
\clearpage
\newpage

\begin{table}[htbp]
\footnotesize
\caption{Constraint equations and target functions. In case 6, X can
stand for either A or B, and $\ph{3}'$ is
defined by eqs.~(\ref{eqn:locnew}) and (\ref{eqn:locmtxn}).\label{tab:0pro}}
\begin{center}
\begin{tabular}{cclcc@{$=$}l}
\\    
\hline
\hline
Case & Units(1-5) &Geometrical Constraints & Classification & \multicolumn{2}{c}{Comments} \\ 
\hline 
\hline
1 & AABAB & $|\rv{4}(\ph{6}) - \rv{2}(\ph{1})|^2 - {l_{2,4}}^2 = 0 $ 
	    & dist   & $\ph{6}$&$f(\ph{1})$  \\ 
  &         & $\uv{3}(\ph{1},\ph{2})\cdot\uv{6}- \cos\thet{4} = 0 $ 
	    & dot1  & $\ph{2}$&$f(\ph{1})$  \\
  &         & $\left| \rv{4}\left[\ph{6}(\ph{1})\right] - 
		\rv{3\mathrm{h}}\left[\ph{2}(\ph{1})\right]\right|^2-
		{l_{3\mathrm{h},4}}^2$ & target & 0&\fph{1}\\ \hline
2 & AAAAB & $| \rv{4}(\ph{6}) - \rv{2}(\ph{1}) |^2 - {l_{2,4}}^2 = 0 $ 
	    & dist   & $\ph{6}$&$f(\ph{1})$  \\ 
  &         & $\left[\rv{5\mathrm{t}}-\rv{3}(\ph{1},\ph{2})\right]\cdot\uv{6}
		-l_{3,4}\cos\thet{4}-l_{4,5\mathrm{h}}-
		l_{5\mathrm{h},5\mathrm{t}}\cos\thet{5\mathrm{t}}= 0 $
	    & dot   & $\ph{2}$&$f(\ph{1})$  \\
  &         & $\left| \rv{4}\left[\ph{6}(\ph{1})\right] - \rv{3}\left[\ph{2}(\ph{1})\right] \right|^2- 
		{l_{3,4}}^2=0 $ & target & $0$&\fph{1}\\ \hline
3 & BABAB & $| \rv{4}(\ph{6}) - \rv{2}(\ph{1}) |^2 - {l_{2,4}}^2 = 0 $ 
	    & dist   & $\ph{6}$&$f(\ph{1})$  \\ 
  &         & $\uv{3}(\ph{2})\cdot\uv{6}-\cos\thet{4} = 0 $ 
	    & quad  & \multicolumn{2}{c}{determine $\ph{2}$}  \\
  &         & $\left| \rv{4}\left[\ph{6}(\ph{1})\right] - 
		\rv{3\mathrm{h}}(\ph{1},\ph{2}) \right|^2-
		{l_{3\mathrm{h},4}}^2 = 0 $ & target & $0$&\fph{1}\\ \hline
4 & BABAA & $| \rv{4}(\ph{6}) - \rv{2}(\ph{1}) |^2 - {l_{2,4}}^2 = 0 $ 
	    & dist   & $\ph{1}$&$f(\ph{6})$  \\ 
  &         & $\uv{4}(\ph{6},\ph{5})\cdot\uv{1} - \cos\thet{2} = 0 $ 
	    & dot1  & $\ph{5}$&$f(\ph{6})$  \\
  &         & $| \rv{2}\left[\ph{1}(\ph{6})\right] - \rv{3\mathrm{t}}
		\left[\ph{5}(\ph{6})\right] |^2-
		{l_{2,3\mathrm{t}}}^2 = 0 $ & target & $0$&\fph{6}\\ \hline
5 & BAAAA & $| \rv{4}(\ph{6}) - \rv{2}(\ph{1}) |^2 - {l_{2,4}}^2 = 0 $ 
	    & dist   & $\ph{1}$&$f(\ph{6})$  \\ 
  &         & $\left[\rv{1\mathrm{h}} - \rv{3}(\ph{6},\ph{5})\right]\cdot\uv{1} + 
		l_{2,3}\cos\thet{2}+l_{1\mathrm{h},1\mathrm{t}}\cos\thet{1\mathrm{h}}
		+l_{1\mathrm{t},2\mathrm{h}}= 0 $ 
	    & dot   & $\ph{5}$&$f(\ph{6})$  \\
  &         & $\left| \rv{2}\left[\ph{1}(\ph{6})\right] - 
		\rv{3}\left[\ph{5}(\ph{6})\right] \right|^2- 
		{l_{2,3}}^2 = 0 $ & target & $0$&\fph{6}\\ \hline 
6 & AXAXA & $| \rv{3}(\ph{3}') - \rv{2\mathrm{h}}(\ph{1}) |^2 
		- {l_{2\mathrm{h},3}}^2=0 $ 
	    & dist      & $\ph{1}$&$f(\ph{3}')$  \\
  &        & $| \rv{3}(\ph{3}') - \rv{4\mathrm{t}}(\ph{6})|^2 
		- {l_{3,4\mathrm{t}}}^2=0$
	    & dist    & $\ph{6}$&$f(\ph{3}')$   \\
  &        & $\left|\rv{2\mathrm{t}}\left[\ph{1}(\ph{3}'),\ph{3}'\right] 
		- \rv{4\mathrm{h}}\left[\ph{6}(\ph{3}'),
	       \ph{3}'\right]\right|^2 - {l_{2\mathrm{t},4\mathrm{h}}}^2 = 0 $ 
	    & target & $0$&$f(\ph{3}')$ \\ \hline
\end{tabular}
\end{center}

\end{table}

%\clearpage
%\newpage

\begin{table}[htbp]
\footnotesize
\caption{The mathematical form of the constraint functions.  Here $i$,
$j$, and $k$ are labels to units, and $d$ is a constant that varies
from case to case.  The reference units of $i$ and $j$ are $i'$ and
$j'$, respectively.  The reference unit of $j'$ is $j''$. The labels
$ a$ and $ b$ are either h or t, depending on whether
the notation is forward or backward.  We define
$ a^*=\mathrm{h}$ when $ a=\mathrm{t}$, and
$ a^*=\mathrm{t}$ when $ a=\mathrm{h}$. The torsional
variables that appear in the constraint equations are \ph{\mathrm{I}}
and \ph{\mathrm{II}}. The constant characteristic matrix is \Mx{}. The
last row defines notation used in this table. Notation not defined
here is defined in Sec.~\ref{sec-RBsch}.
\label{tab:constype}}
\begin{center}
\begin{math}
\begin{array}{lll}
\\
\hline
\hline
\mbox{Type} & \mbox{General function form} & \Mx{} \mbox{  (used in }
\Ph{\mathrm{I}}\Mx{}\Ph{\mathrm{II}}=0\mbox{ )} \\
\hline
\hline
\mbox{quad}     & \uv{i}(\ph{\mathrm{I}})\cdot\uv{j}-d = 0 & \mbox{not
applicable}\\
\hline
\mbox{dist}     & | \rv{i a}(\ph{\mathrm{I}}) - \rv{j b}(\ph{\mathrm{II}})|^2-d=0 
	& (| \rv{i' a}-\rv{j' b}|^2+{l_{i' a,i a}}^2
	+ {l_{j' b,j b}}^2-d)\C  \\
&       & -2\tp{\Lc{i' a}} \tp{\Trx{i'}{lab}}\Trx{j'}{lab}\Lc{j' b} 
+2\tp{\Lc{i' a}} \tp{\Trx{i'}{lab}}\Gm(\rv{i' a}-\rv{j' b}) \\
&       & -2\tp{\left[\Gm(\rv{i' a}-\rv{j' b})\right]}\Trx{j'}{lab}\Lc{j' b} \\
\hline
\mbox{dot}      &
\left[\rv{i a}-\rv{j b}(\ph{\mathrm{I}},\ph{\mathrm{II}})\right]
	\cdot\uv{k} - d = 0 
	& \left[(\rv{i a}-\rv{j'' b})\cdot\uv{k} - d\right]\C  
	  -\tp{\Lc{j'' b}}\tp{\Trx{j''}{lab}}\Gm(\uv{k}) \\
&       & - \G{\uv{k}}{j''}\Trx{j''\theta}{}\Lc{j' b}\\
\hline 
 \mbox{dot1}    & \uv{j}(\ph{\mathrm{I}},\ph{\mathrm{II}})\cdot\uv{k}- d = 0 
	& -d\C + \G{\uv{k}}{j''}\Trx{j''\theta}{}\Lb{j'}  \\  
\hline 
\end{array}
$\\$
\begin{array}{lll}
\C &\equiv &\left( \begin{array}{ccc}
		1       &0      &0 \\
		0       &0      &0 \\
		0       &0      &0 \end{array} \right),\; 
\Gm(\xv{}) \equiv      \left(\begin{array}{ccc}
				\xv{x}  &0      &0 \\
				\xv{y}  &0      &0 \\
				\xv{z}  &0      &0 
				\end{array}
		  \right) \nonumber \\
\G{\xv{}}{i} &\equiv&\left\{ \begin{array}{lll} 
			\left(  \begin{array}{ccc}
				\left[\tp{\Trx{i}{lab}}\xv{}\right]_x   &0      &0 \\

		0&\left[\tp{\Trx{i}{lab}}\xv{}\right]_y&
				\left[\tp{\Trx{i}{lab}}\xv{}\right]_z \\
				0&\left[\tp{\Trx{i}{lab}}\xv{}\right]_z&-\left[\tp{\Trx{i}{lab}}\xv{}\right]_y 
				\end{array} 
			\right) 
					&\mbox{, if} &\thet{i}\neq0 \hspace*{1.2in}\\
			\tp{\left[\Gm(\xv{})\right]}\Trx{i}{lab}
				&\mbox{, if} &\thet{i}=0\hspace*{1.2in}
			\end{array}
		  \right. \nonumber \\
\Lc{i' a}&\equiv        &l_{i'\mathrm{h},i'\mathrm{t}}\Lb{i' a}+
			l_{i' a^*,i a}\Lb{i'} \nonumber \\
{l_{i' a,i a}}^2&       \equiv
	&{l_{i'\mathrm{h},i\mathrm{t}}}^2+{l_{i' a^*,i a}}^2+
	2l_{i'\mathrm{h},i'\mathrm{t}}l_{i' a^*,i a}
	\cos\thet{i'}           \nonumber \\
\hline \nonumber

\end{array} 
\end{math}

\end{center}

\end{table}

%\clearpage
%\newpage

\begin{table}[htbp]
\caption{Comparison of simulation results with different rebridging methods
at 298 K. For all simulations $\Delta\ph{\max}=10^\circ$, except for
WJM, in which $\Delta\ph{\max}=30^\circ$. \label{tab:comp4}}
\begin{center}
\begin{tabular}{cr@{.}lcccc}
\hline
\hline
Method  & \multicolumn{2}{c}{$\Delta\ph{{\mathrm{avg}}}$ (deg)}& $P_\acc$  & $P_{\mathrm{cross}}$ &Number of steps & CPU
time (hrs)\\
\hline
\hline
NJ      &  6&662  &  0.162      & 0.000303 & $4\times10^5$ & 32 \\
WJ      &  7&291  &  0.172      & 0.000414 & $4\times10^5$ & 32 \\
WJO     & 10&006  &  0.177      & 0.000167 & $4\times10^5$ & 34 \\
WJM     & 8&525   &  0.111      & 0.000468 & $2\times10^5$ & 40 \\
MT      &  2&699  &  0.051      & 0.000222 & $4\times10^5$ & 30 \\
\hline
\end{tabular}
\end{center}
\end{table} 
%\clearpage
%\newpage

\begin{table}[htbp]
\begin{center}
\caption{Acceptance probability observed for swapping moves in
the parallel tempering simulation. \label{tab:acc_para}}
\begin{tabular}{r @{$\leftrightarrow$} rc}
\hline
\hline
\multicolumn{2}{c}{Swap}     & $P_\acc$ \\ 
\hline
\hline
298\ K~ & ~500\ K & 0.147 \\
500\ K~ & ~1000\ K & 0.113 \\
1000\ K~ & ~3000\ K & 0.136 \\
\hline
\end{tabular}
\end{center}
\end{table} 
%\clearpage
%\newpage
\renewcommand{\baselinestretch}{1.7}
\begin{table}[htbp]
\caption{Simulation data for tryptophan, lysine,  and arginine
residues of the CNWKRGDC\put(-5,15){\line(-1,0){65}}
\put(-5,10){\line(0,1){5}}
\put(-70,10){\line(0,1){5}}~peptide. \label{tab:residues}} 
\begin{center}
\begin{tabular}{cr@{$\times$}lccr@{.}lccc}
\hline
\hline
Method & $n_1$&$n_2$ & Residue 
&$\Delta\ph{}$/CPU (deg$\cdot$min$^{-1}$)&\multicolumn{2}{c}{$P_\acc$} & CPU (min)& \\
\hline
\hline
SLA      &  1&0   & Trp  & 142    & 0&0317  & 2940 \\
	 &\multicolumn{2}{c}{}& Lys   & 82    & 0&00229 &      \\
	 &\multicolumn{2}{c}{}& Arg   & 90    & 0&00150 &      \\
\hline
SLA      & 10&0   & Trp  & 158   & 0&0846& 1620   \\
	 &\multicolumn{2}{c}{}   & Lys   & 575   & 0&187 &        \\
	 &\multicolumn{2}{c}{}   & Arg  & 152   & 0&0325&        \\
\hline
SLA      & 30&0   & Trp  & 315   & 0&313 & 3910   \\
	 &\multicolumn{2}{c}{}   & Lys   & 526   & 0&214 &        \\
	 &\multicolumn{2}{c}{}   & Arg  & 290   & 0&131 &        \\
\hline
SLA     & 100&0   & Trp   & 153   & 0&373 & 3220   \\
	&\multicolumn{2}{c}{}   & Lys   & 445   & 0&304 &        \\
	&\multicolumn{2}{c}{}   & Arg  & 130   & 0&0821&        \\
\hline
LA 	& 5&5 	& Trp   & 131   & 0&0660& 2040   \\
	&\multicolumn{2}{c}{}   & Lys   & 271   & 0&0947&        \\
	&\multicolumn{2}{c}{}   & Arg  & 157   & 0&0529&        \\
\hline
LA 	&10&5 & Trp   & 147   & 0&0686& 2260   \\
	&\multicolumn{2}{c}{}   & Lys   & 375   & 0&169 &        \\
	&\multicolumn{2}{c}{}   & Arg  & 205   & 0&0792&        \\
\hline
LA	&10&10 & Trp   & 120   & 0&0841& 2800   \\
	&\multicolumn{2}{c}{}   & Lys   & 397   & 0&220 &        \\
	&\multicolumn{2}{c}{}   & Arg  & 225   & 0&118 &        \\
\hline
LA	&20&10	& Trp   & 28  & 0&120 & 2500   \\
	&\multicolumn{2}{c}{}   & Lys   & 355   & 0&348 &        \\
	&\multicolumn{2}{c}{}   & Arg  & 55  & 0&0811&        \\
\hline
LARC    &5&5      & Trp	& 213      & 0&116 &1690    \\
	&\multicolumn{2}{c}{}   & Lys   & 673   & 0&221 &        \\
	&\multicolumn{2}{c}{}   & Arg  & 327   & 0&100 &        \\
\hline
LARC    &10&10     & Trp   & 211   & 0&206 &2560    \\
	&\multicolumn{2}{c}{}   & Lys   & 749   & 0&421 &        \\
	&\multicolumn{2}{c}{}   & Arg  & 248   & 0&146 &        \\
\hline
LARC 	&15 &15	& Trp   & 151   & 0&334 & 4250  \\
	&\multicolumn{2}{c}{}   & Lys   & 548   & 0&497 &       \\
	&\multicolumn{2}{c}{}   & Arg  & 346   & 0&357 &       \\
\hline
LARC 	&20&20	& Trp   & 104  	& 0&415 & 6420	\\
	&\multicolumn{2}{c}{}   & Lys   & 394  	& 0&571 &       \\
	&\multicolumn{2}{c}{}   & Arg  & 212  	& 0&350 &       \\
\hline
\end{tabular}

\end{center}
\end{table}
\renewcommand{\baselinestretch}{1.0}

\begin{figure}[p]
\caption{A typical cyclic peptide,
CNWKRGDC\put(-5,15){\line(-1,0){65}}
\put(-5,10){\line(0,1){5}}
\put(-70,10){\line(0,1){5}}. The disulfide bond is shown at the left,
spanned by two (spherical) sulfur atoms. Note that this is not the
lowest free energy conformation of this molecule.\label{fig:CNWKRGDC}}
\end{figure}
\begin{figure}[p]
\caption{Sketches for units of class A and B. The angle $\theta$ is zero for
class B. Only backbone atoms are depicted here. Bold lines indicates
bonds that do not rotate.
\label{fig:clsunit}}
\end{figure}
\begin{figure}[p]
\caption{A segment selected to be rebridged. Change of
driver angles \ph{0} and \ph{7} breaks the connectivity. The dotted area
represents the region in which the positions of the backbone atoms must
be restored.\label{fig:babaa}}
\end{figure}
\begin{figure}[p]
\caption{The rebridging method applied to an ABABA segment.  It can be seen
that $|\rv{1}-\rv{3}|$ and $|\rv{3}-\rv{5}|$ are
constants. The dotted area represents the
region in which the positions of backbone atoms are to be
restored.\label{fig:ababa}}
\end{figure}
\begin{figure}[p]
\caption{A typical target function. Only the $\ph{3}'$ domain where the
target function exists is shown here. The number of branches is four.
\label{fig:2-2var}}
\end{figure}
\begin{figure}[p]
\caption{Shown as solid is the definition of atom groups used for side
chain regrowth:  a) in non-look-ahead and b) in
semi-look-ahead. The left two bonds are connected to other 
rigid units.}
\label{fig:sidechai}
\end{figure}
\begin{figure}[p]
\caption{Schematic pictures for generating trial moves.  The
solid circle 0 denotes an existing unit.  Dotted circles represent trial
configurations of the next unit.  a) CBMC without look-ahead generates and
regrows one unit at one time.  Configuration 2 has the lowest energy and is
most likely to be picked.  b) In look-ahead, we generate
two units and regrow one unit. The configurations generated from 2
turn out to be disfavored, and configuration 4 is chosen instead.
\label{fig:lookahead}}
\end{figure}
\begin{figure}[p]
\caption{The probability distribution for the $\mathrm{C_\beta SSC_\beta}$
disulfide torsional angle observed in NJ, WJ, WJO, WJM, and MT.}
\label{fig:CSSC}
\end{figure}
\begin{figure}[p]
\caption{The energy histograms for all the systems in the parallel
tempering simulation.}
\label{fig:e_histo}
\end{figure}
\begin{figure}[p]
\caption{The histograms for the $\mathrm{C_\beta SSC_\beta}$
disulfide torsional angle observed in the parallel tempering
simulation. The total number of Monte Carlo cycles is N.
The distribution at 3000 K is shown for comparison.}
\label{fig:csscpara}
\end{figure}
\begin{figure}[p]
\caption{Equilibration of the side chains of CNWKRGDC
\put(-5,15){\line(-1,0){65}}
\put(-5,10){\line(0,1){5}}
\put(-70,10){\line(0,1){5}}.  The
numbers at the upper right are the values of $n_1$.}
\label{fig:old_and_new}
\end{figure}
\begin{center}
\clearpage
\newpage
\epsfig{file=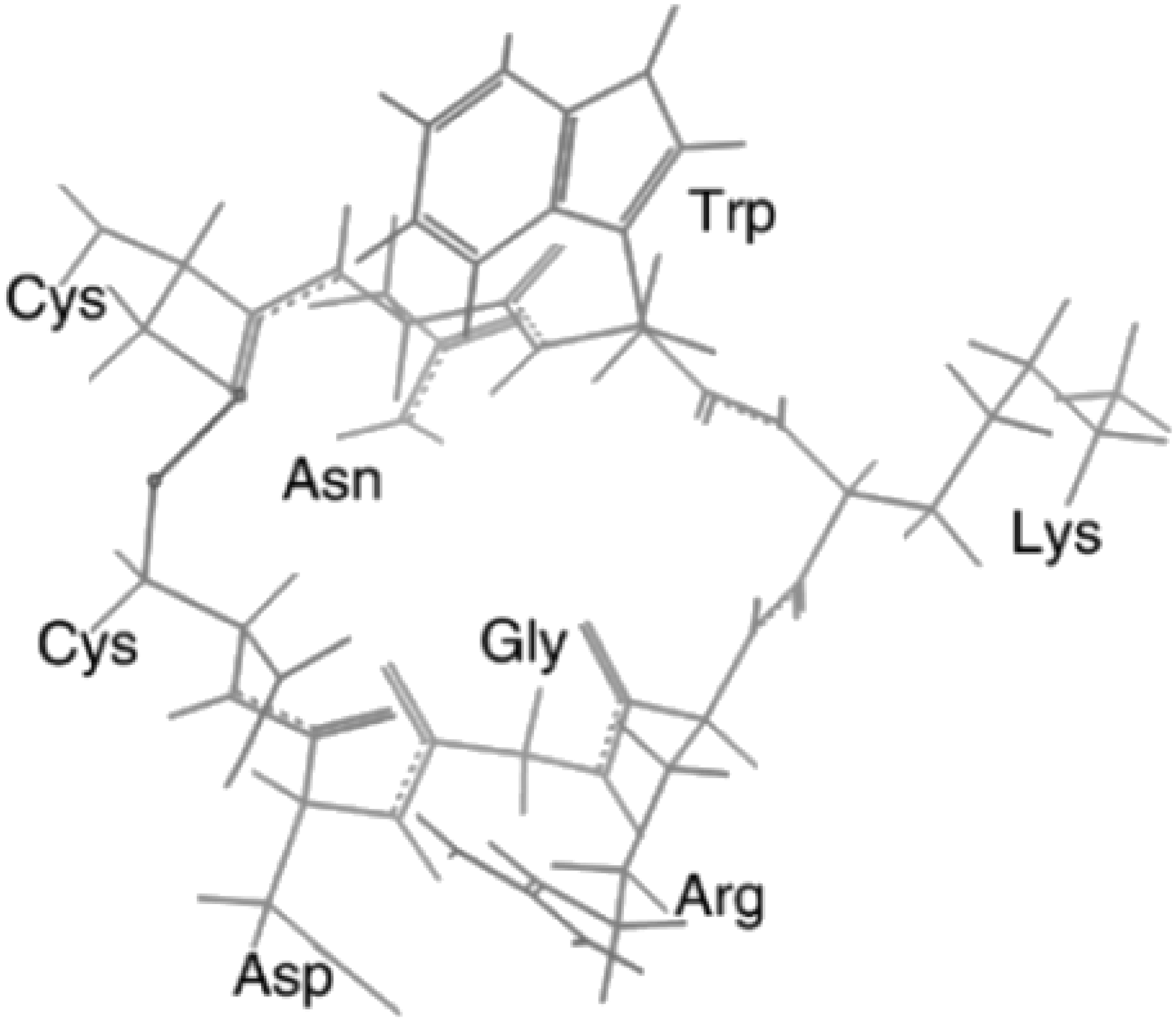,width=5.5in}\\[.5in]
{Figure \ref{fig:CNWKRGDC}: Wu and Deem, `Efficient Monte Carlo\ldots'}
\clearpage
\newpage
\epsfig{file=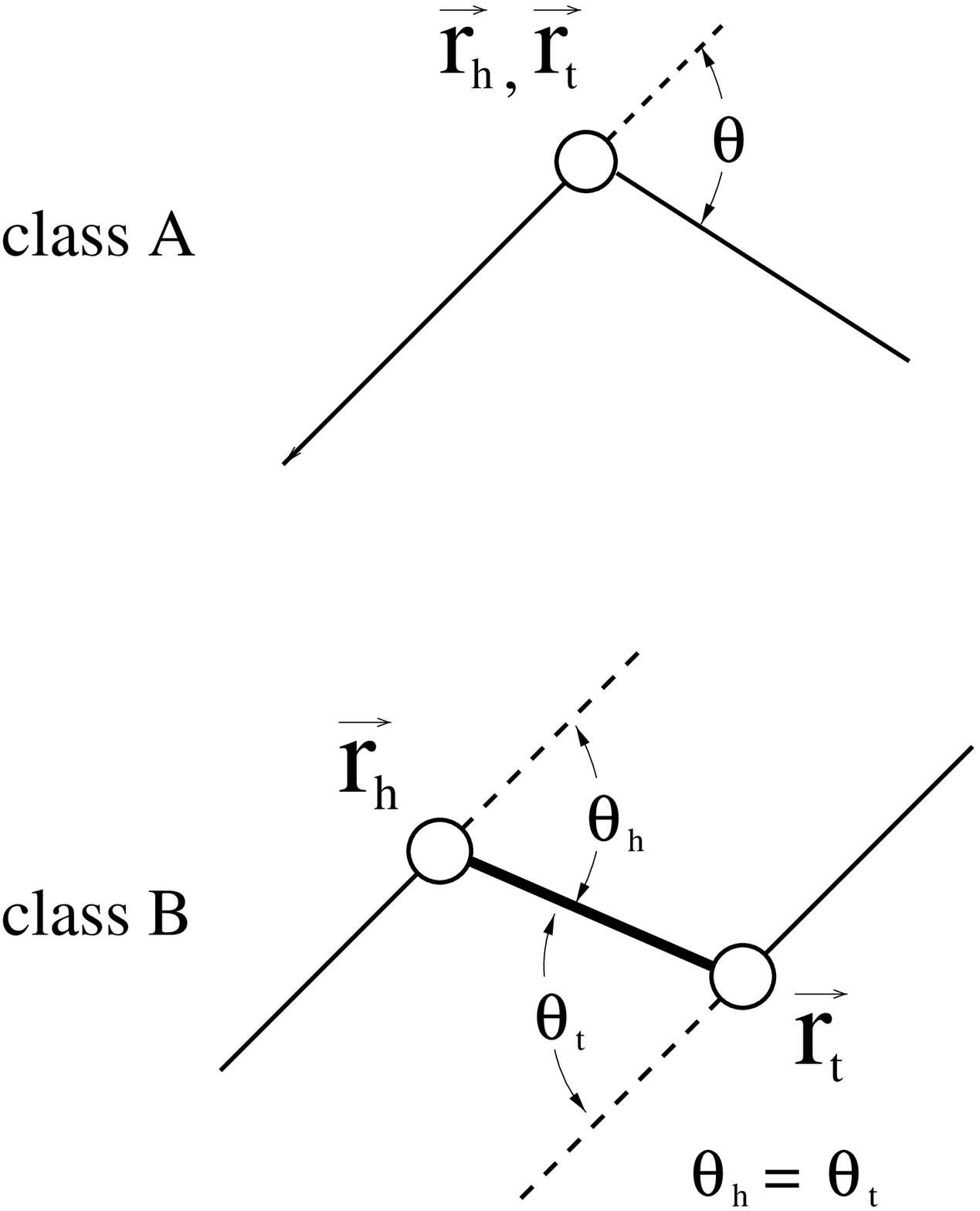,width=4in}\\[.5in]
{Figure \ref{fig:clsunit}: Wu and Deem, `Efficient Monte Carlo\ldots'}
\clearpage
\newpage
\epsfig{file=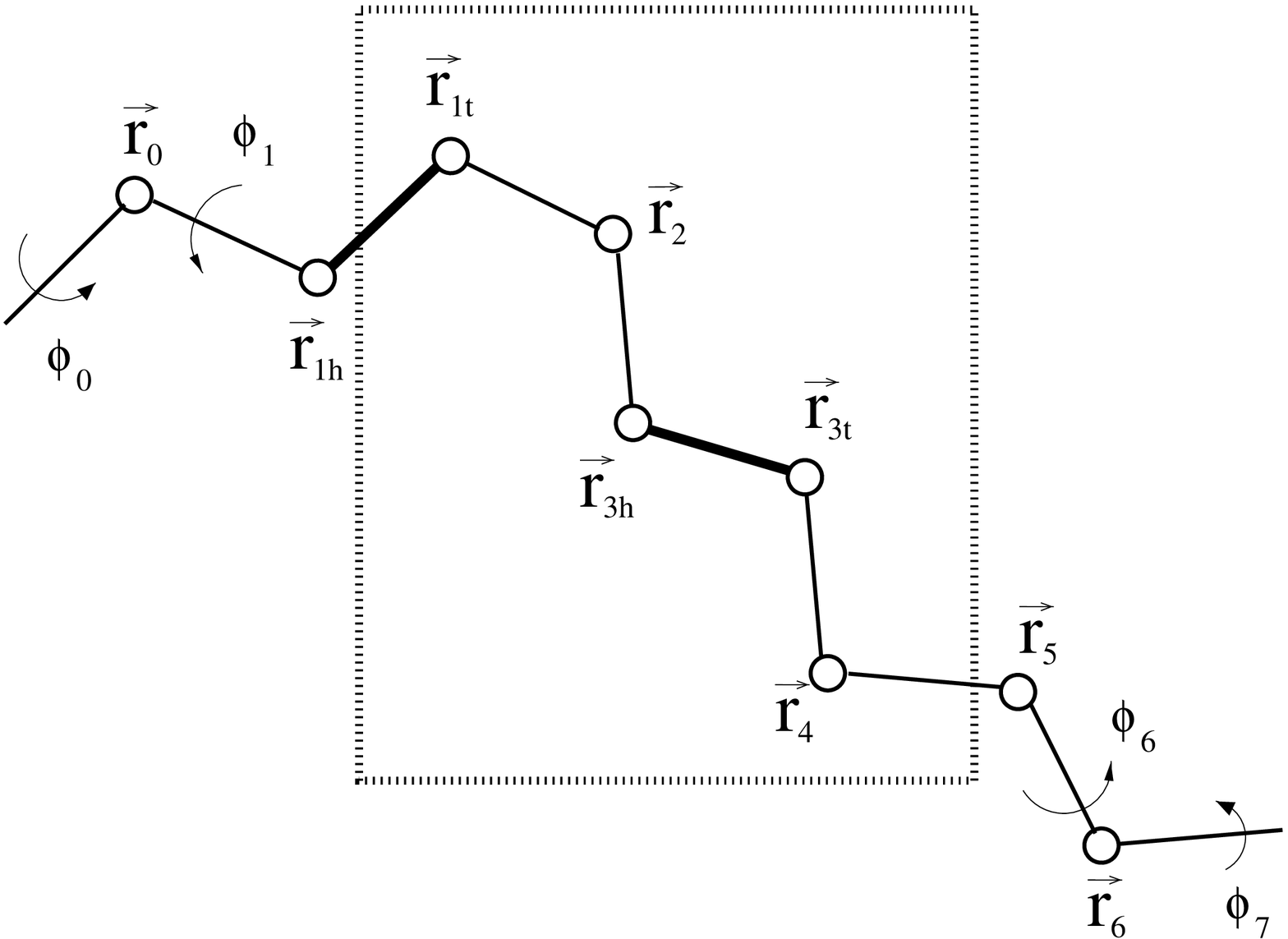,width=6in}\\[.5in]
{Figure \ref{fig:babaa}: Wu and Deem, `Efficient Monte Carlo\ldots'}
\clearpage
\newpage
\epsfig{file=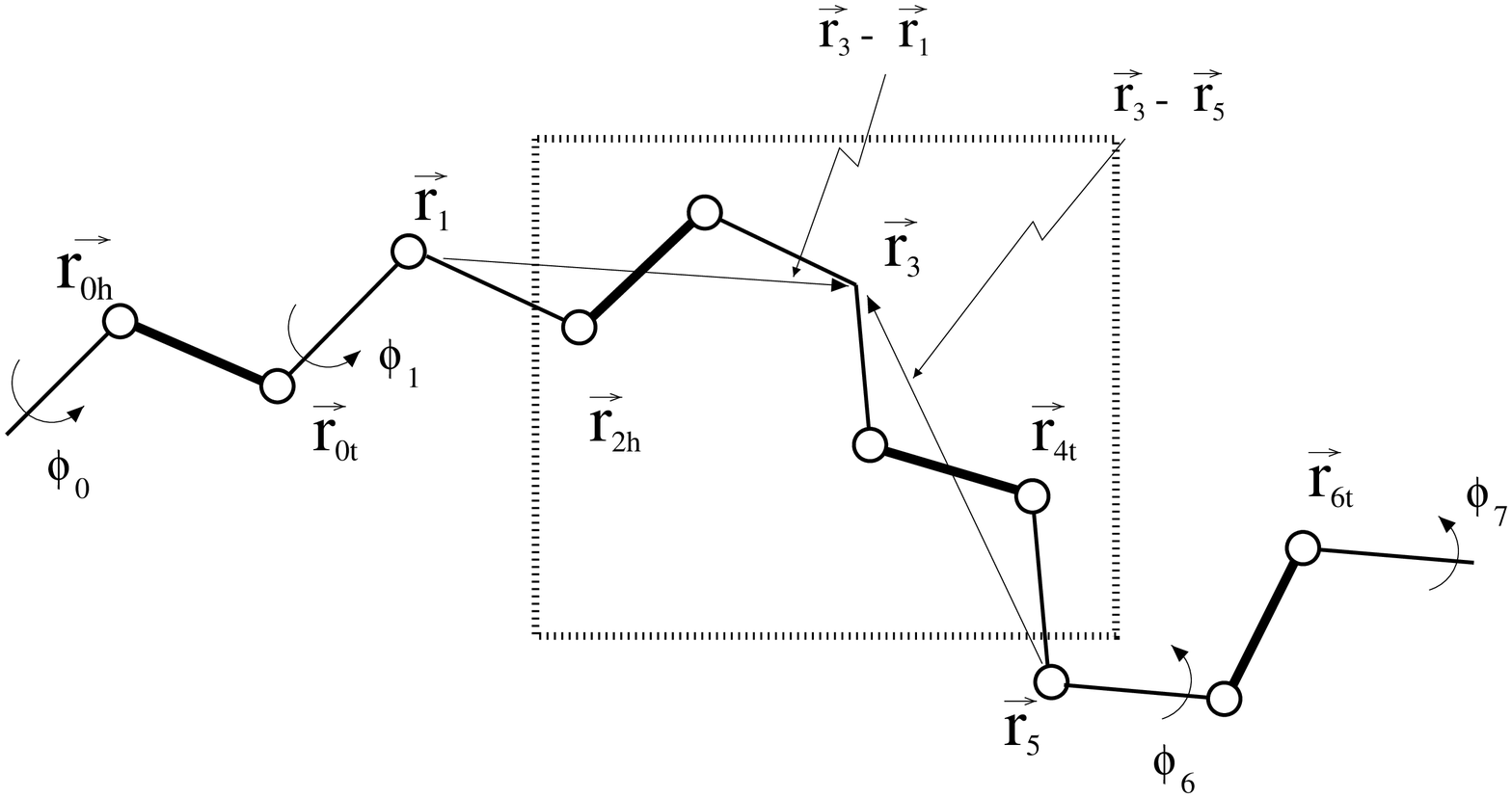,width=6in}\\[.5in]
{Figure \ref{fig:ababa}: Wu and Deem, `Efficient Monte Carlo\ldots'}
\clearpage
\newpage
\epsfig{file=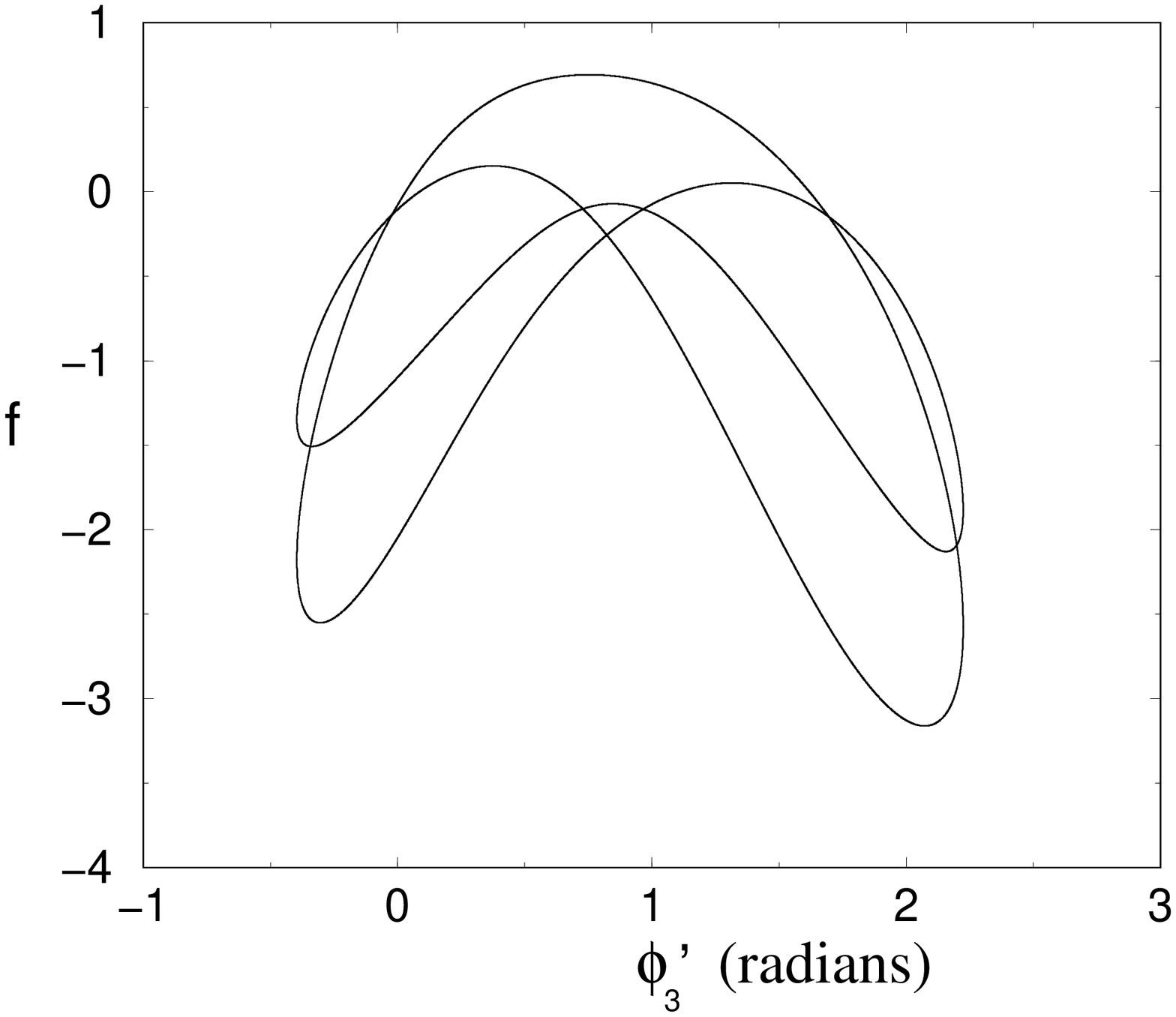,width=5in,angle=270}\\[.5in]
{Figure \ref{fig:2-2var}: Wu and Deem, `Efficient Monte Carlo\ldots'}
\clearpage
\newpage
\epsfig{file=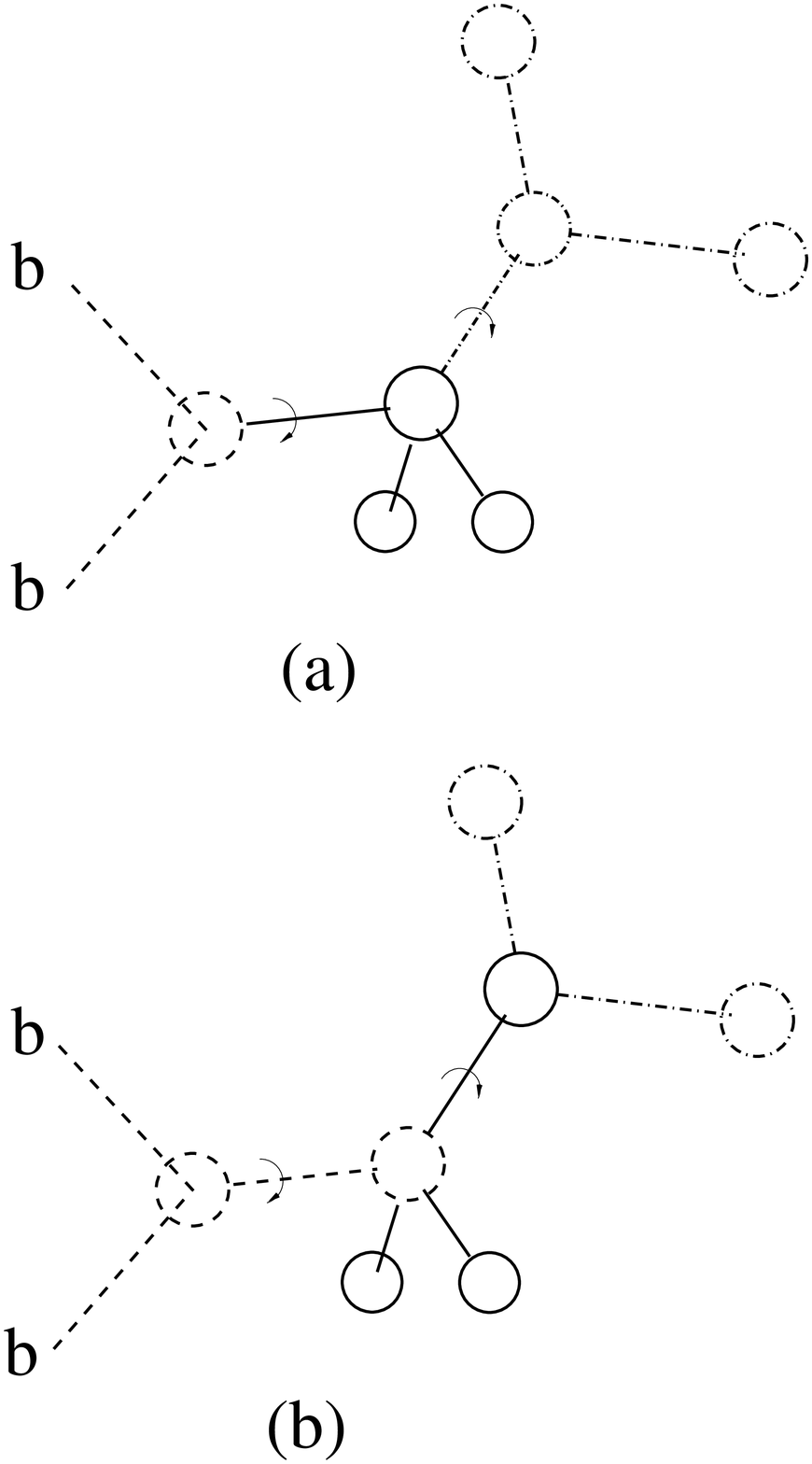,width=3.5in}\\[.5in]
{Figure \ref{fig:sidechai}: Wu and Deem, `Efficient Monte Carlo\ldots'}
\clearpage
\newpage
\epsfig{file=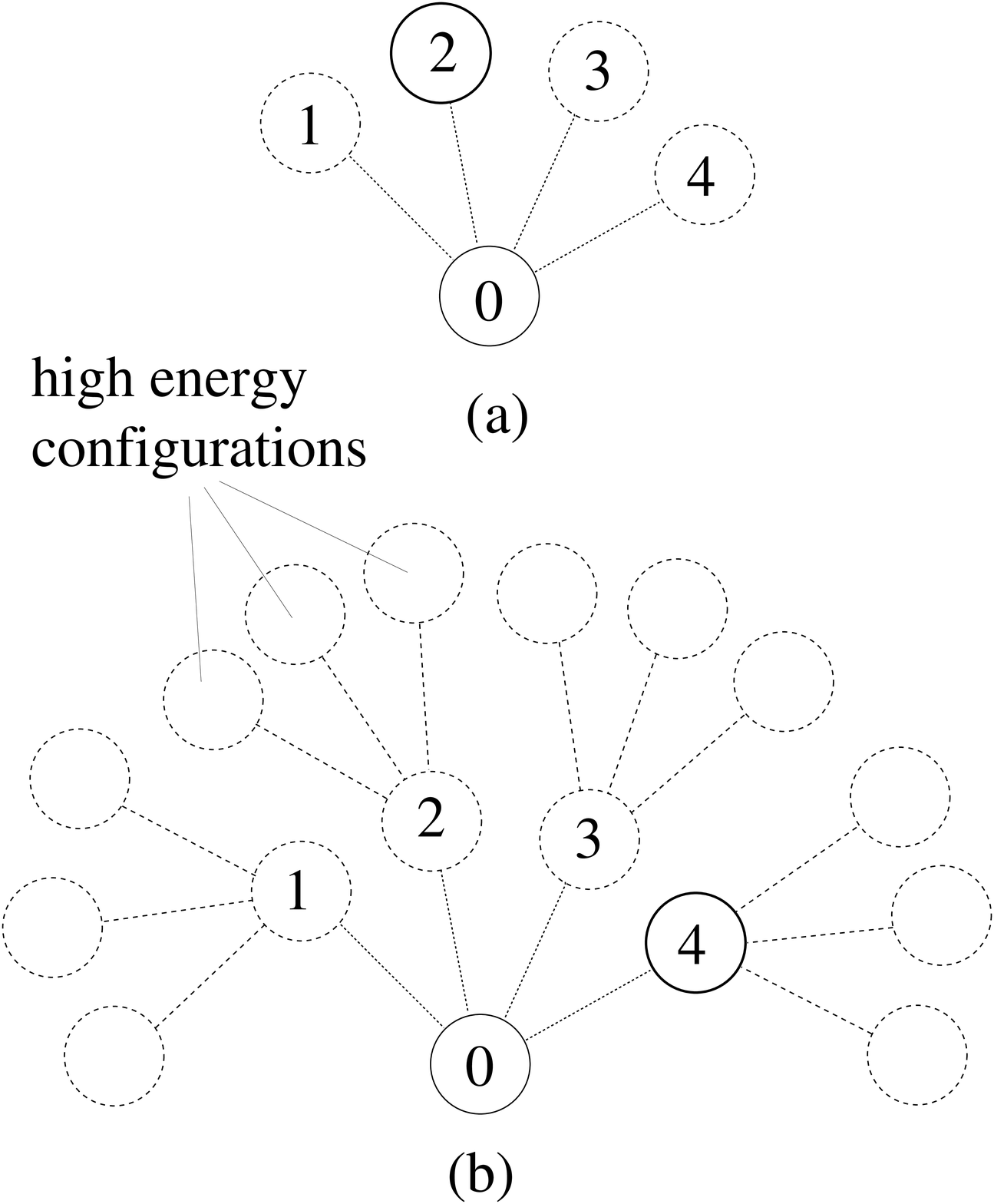,width=4in}\\[.5in]
{Figure \ref{fig:lookahead}: Wu and Deem, `Efficient Monte Carlo\ldots'}
\clearpage
\newpage
\epsfig{file=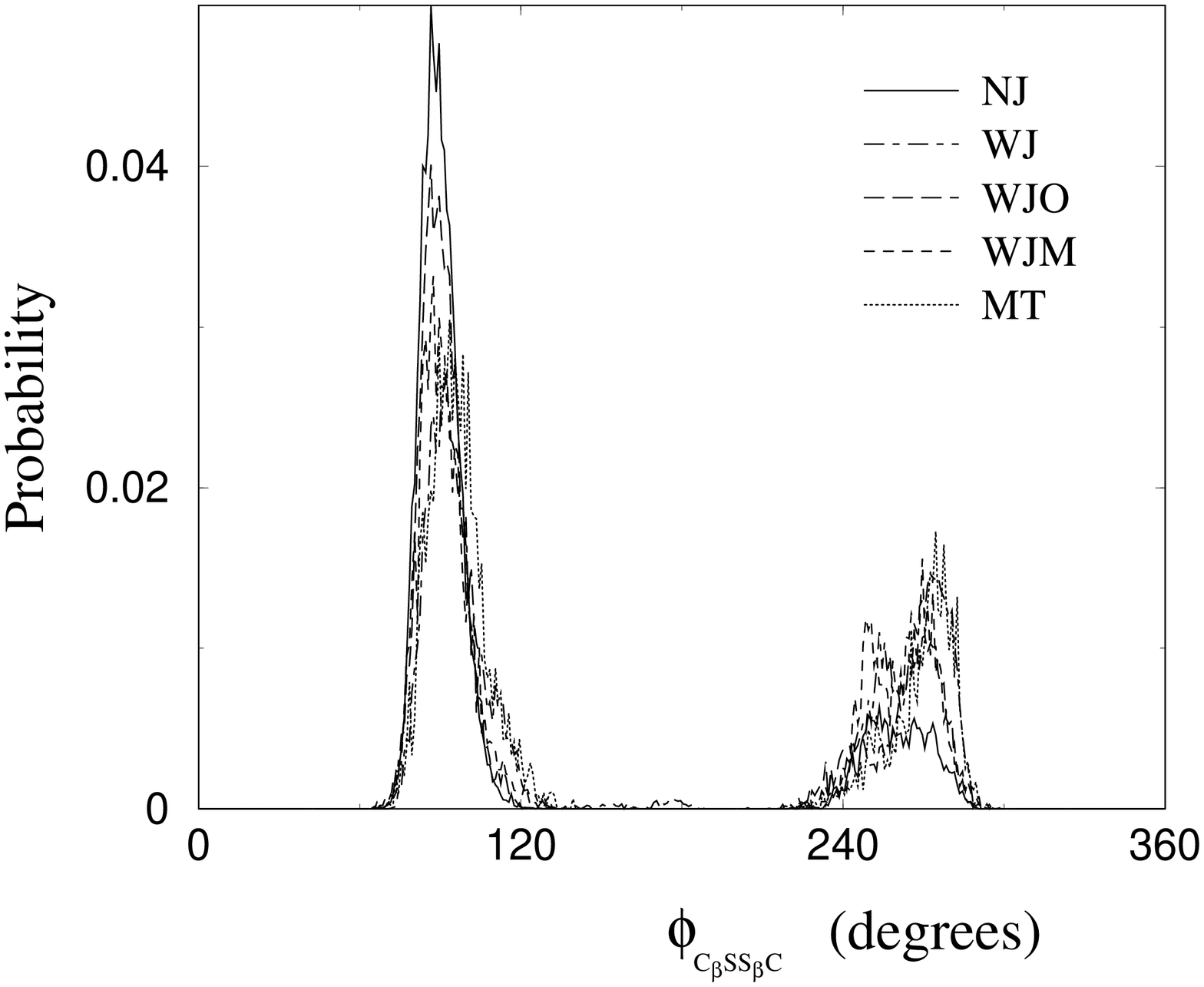,width=5in,angle=270}\\[.5in]
{Figure \ref{fig:CSSC}: Wu and Deem, `Efficient Monte Carlo\ldots'}
\clearpage
\newpage
\epsfig{file=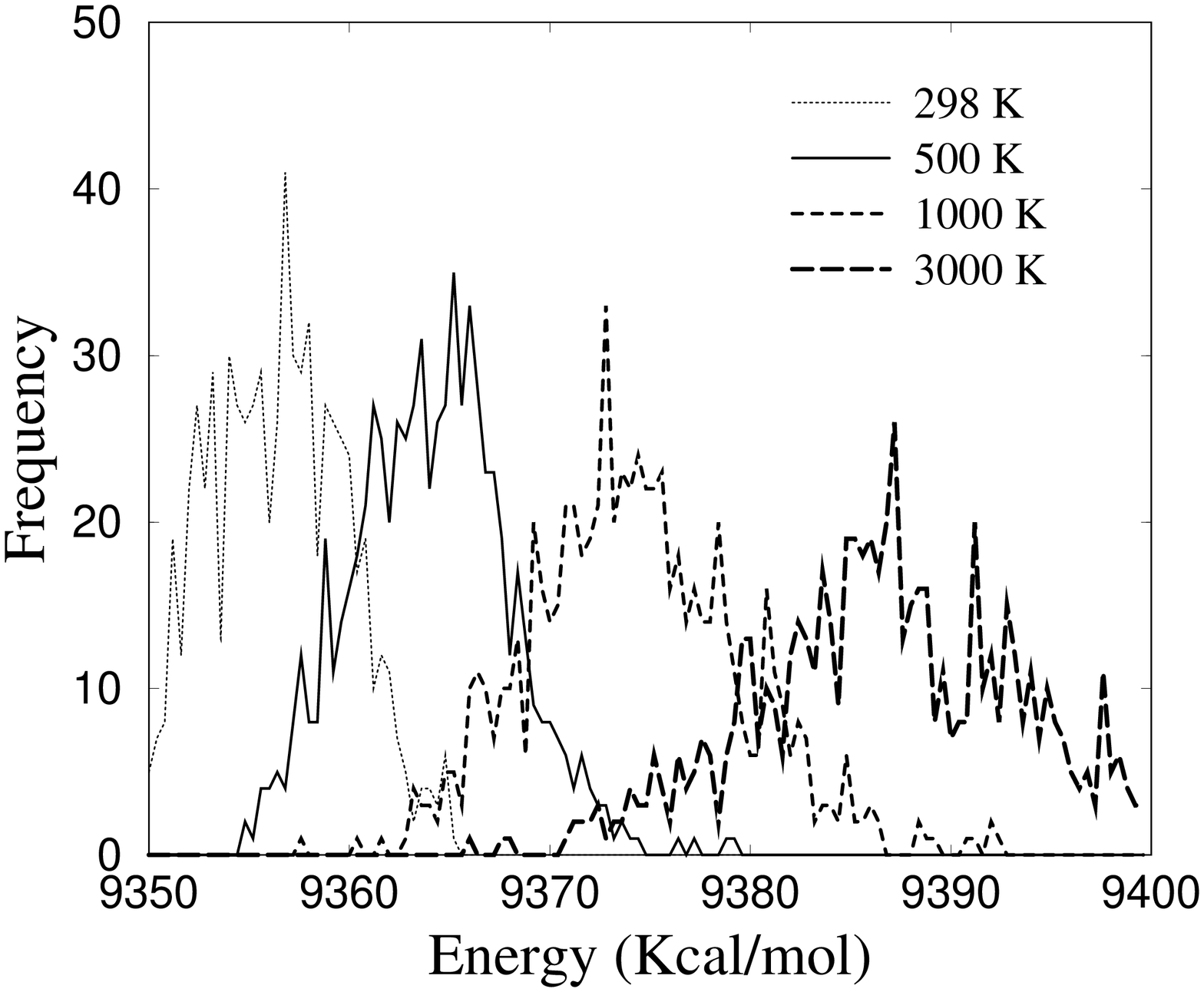,width=5in, angle=270}\\[.5in]
{Figure \ref{fig:e_histo}: Wu and Deem, `Efficient Monte Carlo\ldots'}
\clearpage
\newpage
\epsfig{file=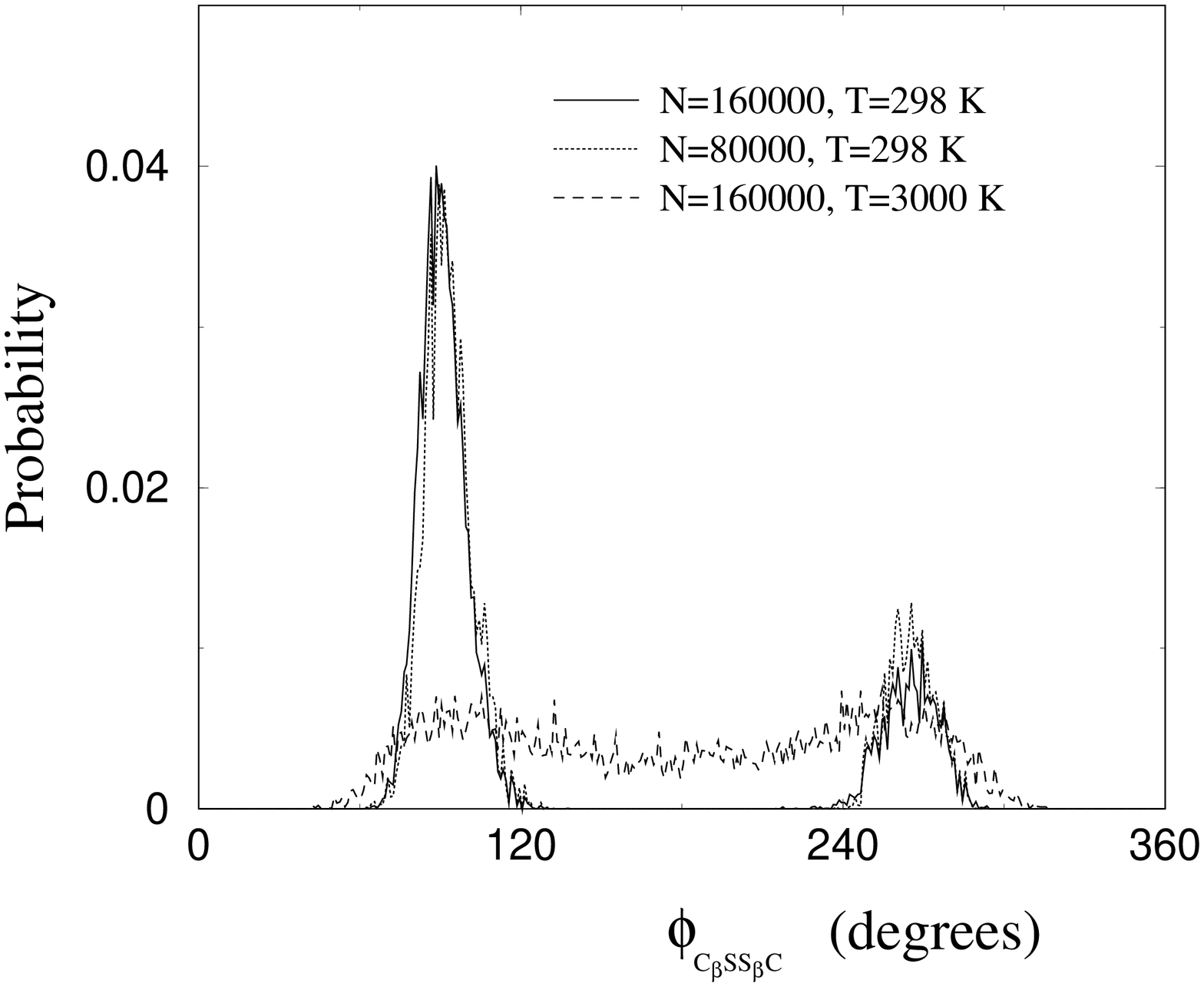,width=5in,angle=270}\\[.5in]
{Figure \ref{fig:csscpara}: Wu and Deem, `Efficient Monte Carlo\ldots'}
\clearpage
\newpage
\epsfig{file=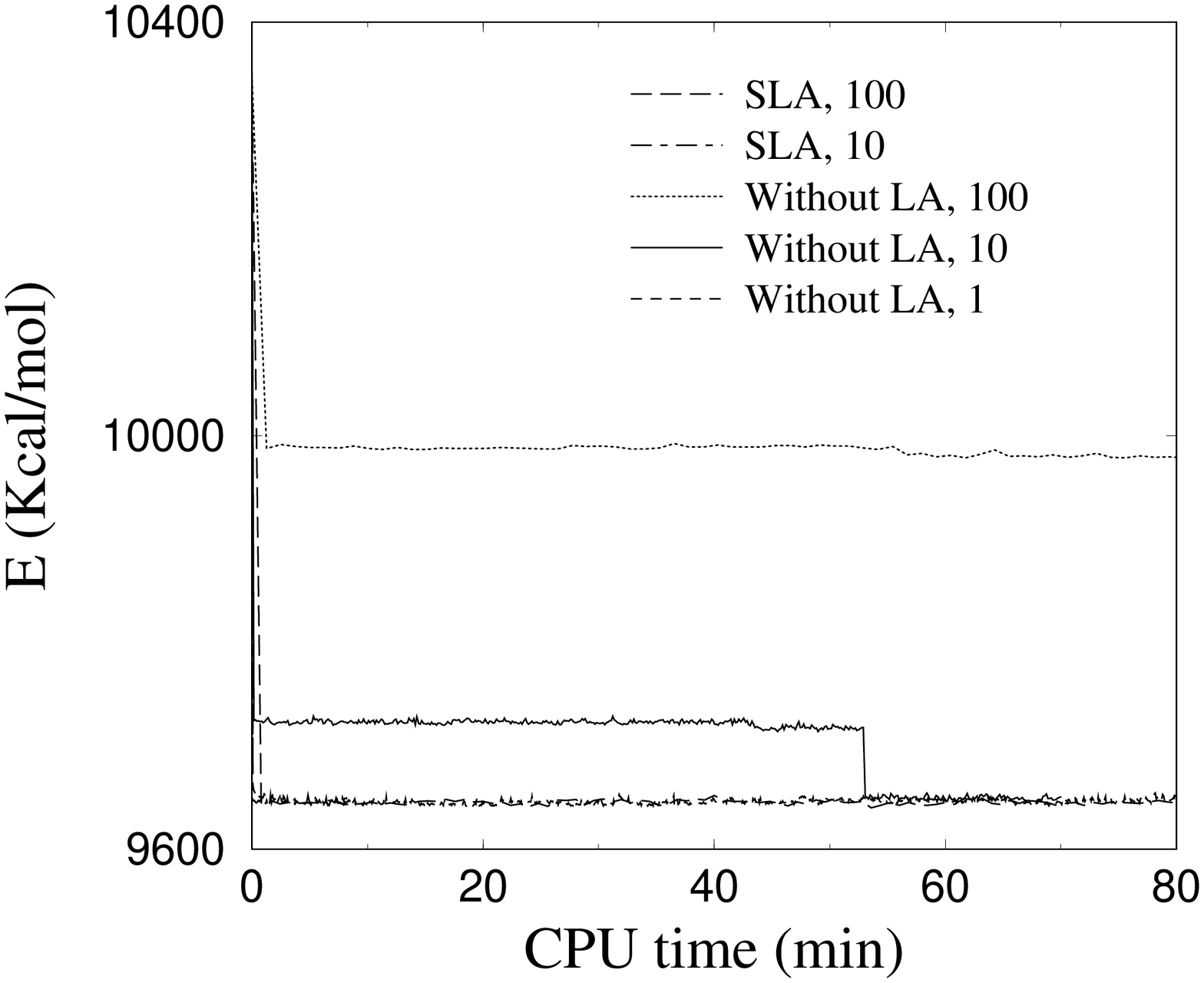,width=5in,angle=270}\\[.5in]
{Figure \ref{fig:old_and_new}: and Deem, `Efficient Monte Carlo\ldots'}
\end{center}

\end{document}